\documentclass[12pt,draftclsnofoot,journal,onecolumn]{IEEEtran}

\usepackage{graphicx}
\usepackage[cmex10]{amsmath}
\usepackage{amsfonts}
\usepackage{amssymb}
\usepackage{mathrsfs}
\usepackage{bm}
\usepackage{algorithmic}
\usepackage{algorithm}
\usepackage{array}
\usepackage{mdwmath}
\usepackage{mdwtab}
\usepackage[caption=false,font=footnotesize]{subfig}
\usepackage{fixltx2e}
\usepackage{stfloats}
\usepackage{footnote}
\usepackage{tabularx}
\usepackage{multirow}
\usepackage{color}
\usepackage{textcomp}
\usepackage{subfig}

\newtheorem{theorem}{Theorem}
\newtheorem{lemma}{Lemma}

\begin{document}

\title{Foresighted Demand Side Management}

\author{Yuanzhang~Xiao and~Mihaela~van~der~Schaar,~\IEEEmembership{Fellow,~IEEE} \\
Department of Electrical Engineering, UCLA. \{yxiao,mihaela\}@ee.ucla.edu.}

\maketitle

\vspace{-1.5cm}
\begin{abstract}
We consider a smart grid with an independent system operator (ISO), and distributed aggregators who have energy storage and purchase energy from the
ISO to serve its customers. All the entities in the system are \emph{foresighted}: each aggregator seeks to minimize its own \emph{long-term}
payments for energy purchase and operational costs of energy storage by deciding how much energy to buy from the ISO, and the ISO seeks to minimize
the \emph{long-term} total cost of the system (e.g. energy generation costs and the aggregators' costs) by dispatching the energy production among
the generators. The decision making of the foresighted entities is complicated for two reasons. First, the information is decentralized among the
entities: the ISO does not know the aggregators' states (i.e. their energy consumption requests from customers and the amount of energy in their
storage), and each aggregator does not know the other aggregators' states or the ISO's state (i.e. the energy generation costs and the status of the
transmission lines). Second, the coupling among the aggregators is unknown to them due to their limited information. Specifically, each aggregator's
energy purchase affects the price, and hence the payments of the other aggregators. However, none of them knows how its decision influences the price
because the price is determined by the ISO based on its state. We propose a design framework in which the ISO provides each aggregator with a
conjectured future price, and each aggregator distributively minimizes its own long-term cost based on its conjectured price as well as its
locally-available information. The proposed framework can achieve the social optimum despite being decentralized and involving complex coupling among
the various entities interacting in the system.
Simulation results show that the proposed foresighted demand side management achieves significant reduction in the total cost, compared to the
optimal myopic demand side management (up to 60\% reduction), and the foresighted demand side management based on the Lyapunov optimization framework
(up to 30\% reduction).
\end{abstract}


\section{Introduction}
The power systems are undergoing drastic changes on both the supply side and the demand side. On the supply side, an increasing amount of renewable
energy (e.g. wind energy, solar energy) is penetrating the power systems. The adoption of renewable energy reduces the environmental damage caused by
conventional energy generation, but also introduces high fluctuation and uncertainty in energy generation on the supply side. To cope with this
uncertainty in energy generation, the demand side is deploying various solutions, one of which is the use of energy storage \cite{ChandyLowTopcuXu}.
In this paper, we study the optimal demand side management (DSM) strategy in the presence of energy storage, and the corresponding optimal economic
dispatch strategy.

Specifically, we consider a power system consisting of energy generators on the supply side, an independent system operator (ISO) that operates the
system, and multiple aggregators and their customers on the demand side. On the supply side, the ISO receives energy purchase requests from the
aggregators as well as reports of (parameterized) energy generation cost functions from the generators, and based on these, dispatches the energy
generators and determines the unit energy prices. On the demand side, the aggregators are located in different geographical areas and provide energy
for residential customers (e.g. households) or for commercial customers (e.g. an office building) in the neighborhood. In the literature, the term
``DSM'' has been broadly used for different decision problems on the demand side. For example, some papers (see
\cite{Mohsenian-Rad_TransSmartGrid}--\cite{KimRenVDS_JSAC2013} for representative papers) focus on the interaction between one aggregator and its
customers, and refer to DSM as determining the power consumption schedules of the users. Some papers
\cite{MalekianOzdaglarWei}--\cite{HuangWalrand_QoU_PES} focus on how multiple aggregators \cite{MalekianOzdaglarWei}--\cite{ScutariPalomar2013b} or a
single aggregator \cite{Tong_DR_Allerton2012}--\cite{HuangWalrand_QoU_PES} purchases energy from the ISO based on the energy consumption requests
from their customers. Our paper pertains to the second category of research works.

The key feature that sets apart our paper from most existing works \cite{MalekianOzdaglarWei}--\cite{ScutariPalomar2013b} is that all the decision
makers in the system are \emph{foresighted}. Each aggregator seeks to minimize its \emph{long-term} cost, consisting of its operational cost of
energy storage and its payment for energy purchase. In contrast, in most existing works \cite{MalekianOzdaglarWei}--\cite{ScutariPalomar2013b}, the
aggregators are \emph{myopic} and seek to minimizing their \emph{short-term} (e.g. one-day or even hourly) cost. In the presence of energy storage,
foresighted DSM strategies can achieve much lower costs than myopic DSM strategies because the current decisions of the aggregators will affect their
future costs. For example, an aggregator can purchase more energy from the ISO than that requested from its customers, and store the unused energy in
the energy storage for future use, if it anticipates that the future energy price will be high. Hence, the current purchase from the aggregators will
affect how much they will purchase in the future. In this case, it is optimal for the entities to make \emph{foresighted} decisions, taking into
account the impact of their current decisions on the future. Since the aggregators deploy foresighted DSM strategies, it is also optimal for the ISO
to make foresighted economic dispatch, in order to minimize the \emph{long-term} total cost of the system, consisting of the long-term cost of energy
generation and the aggregators' long-term operational cost. Note that although some works \cite{Tong_DR_Allerton2012}--\cite{HuangWalrand_QoU_PES}
assume that the aggregator is foresighted, they study the decision problem of a \emph{single} aggregator and do not consider the economic dispatch
problem of the ISO. When there are multiple aggregators in the system (which is the case in practice), this approach neglects the impact of
aggregators' decisions on each other, which leads to suboptimal solutions in terms of minimizing the total cost of the system.

When the ISO and \emph{multiple} aggregators make \emph{foresighted} decisions, it is difficult to obtain the optimal foresighted strategies for two
reasons. First, the information is decentralized. The total cost depends on the generation cost functions (e.g. the speed of wind for wind energy
generation, the amount of sunshine for solar energy generation, and so on), the status of the transmission lines (e.g. the flow capacity of the
transmission lines), the amount of electricity in the energy storage, and the demand from the customers, all of which may change due to supply and
demand uncertainty. However, none of the entities knows all the above information: the ISO knows only the generation cost functions and the status of
the transmission lines, and each aggregator knows only the status of its own energy storage and the demand from its own customers. Hence, the DSM
strategy needs to be decentralized, such that each entity can make decisions solely based on its locally-available information. Second, the
aggregators are coupled in a complicated way that is unknown to them. Specifically, each aggregator's purchase affects the prices, and thus the
payments of the other aggregators. However, the price is determined by the ISO based on the generation cost functions and the status of the
transmission lines, neither of which is known to any aggregator. Hence, each aggregator does not know how its purchase will influence the price,
which makes it difficult for the aggregator to make the optimal decision.

To overcome the difficulty resulting from information decentralization and complicated coupling, we propose a decentralized DSM strategy based on
conjectured prices. Specifically, each aggregator makes decisions based on its conjectured price, and its local information on the status of its
energy storage and the demand from its customers. In other words, each aggregator summarizes all the unavailable information into its conjectured
price. Note, however, that the price is determined based on the generation cost functions and the status of the transmission lines, which is only
known to the ISO. Hence, the aggregators' conjectured prices are determined by the ISO. We propose a simple online algorithm for the ISO to update
the conjectured prices based on its local information, and prove that by using the algorithm, the ISO obtains the optimal conjectured prices under
which the aggregators' (foresighted) best responses minimize the total cost of the system.


\begin{table}
\renewcommand{\arraystretch}{1.0}
\caption{Comparisons With Related Works on Demand-Side Management.} \label{table:RelatedWork} \centering
\begin{tabular}{|c|c|c|c|c|c|c|}
\hline
& Energy storage & Time horizon & Foresighted & Aggregators & Supply uncertainty & Demand Uncertainty \\
\hline
\cite{Mohsenian-Rad_TransSmartGrid}\cite{LiChenLow}\cite{MalekianOzdaglarWei} & No & 1 day & No  & Multiple & No & No \\
\hline
\cite{JiangLow} & No & 1 day  & No & Multiple & Yes & No \\
\hline
\cite{KimRenVDS_JSAC2013} & No & 1 day  & No & Multiple & No & Yes \\
\hline
\cite{ChandyLowTopcuXu} & Yes & 1 day  & No & Multiple & No & No \\
\hline
\cite{ScutariPalomar2013a}\cite{ScutariPalomar2013b} & Yes & 1 day  & No & Multiple & Yes & Yes \\
\hline
\cite{Tong_DR_Allerton2012}\cite{Tong_Allerton2013} & Yes & Infinite & Yes  & Single & No & Yes \\
\hline
\cite{HuangWalrand_EnergyStorage_TR}\cite{HuangWalrand_QoU_PES} & Yes & Infinite  & Yes & Single & Yes & Yes \\
\hline
Proposed & Yes & Infinite  & Yes & Multiple & Yes & Yes \\
\hline
\end{tabular}
\end{table}

\subsection{Related Works on Demand-Side Management}
The early works \cite{Mohsenian-Rad_TransSmartGrid}--\cite{MalekianOzdaglarWei} on demand side management focused on the power consumption scheduling
of the customers in one day. They considered a power system with no demand uncertainty (e.g. fixed, instead of stochastic, demand) and no supply
uncertainty (e.g. no renewable energy generation). In this simplified model, they formulated the cost minimization problem
\cite{Mohsenian-Rad_TransSmartGrid} or the utility maximization problem \cite{LiChenLow}\cite{MalekianOzdaglarWei} as a convex optimization problem,
and proposed distributed algorithms to achieve the optimal power consumption scheduling.

However, with the penetration of renewable energy, there is a high degree of uncertainty in the supply. In addition, the demand cannot be the same
for every day. Hence, we need to build an appropriate model for the power system that takes into account the uncertainty in the supply and demand.
Some works \cite{JiangLow} considered such a model with demand and supply uncertainties, but still focused on the decision problem in one day. Since
the users are myopic and optimize its one-day utility, the problem is formulated as a finite-horizon Markov decision problem (MDP) and a greedy
algorithm is proposed to achieve the optimal solution  \cite{JiangLow}.

Nevertheless, the above works neglected an important trend in the future power system: the increasing adoption of energy storage on the demand side
to cope with the uncertain energy supply caused by renewable energy generation. Some works \cite{ChandyLowTopcuXu}--\cite{ScutariPalomar2013b}
propose myopic DSM strategies to minimize the short-term (e.g. one-day or hourly) cost in the presence of energy storage. However, with energy
storage, the strategies that minimize the long-term cost can greatly outperform the myopic strategies that minimize the short-term cost. For example,
in a myopic strategy, the aggregator may tend to purchase as little power as possible as long as the demand is fulfilled, in order to minimize the
current operational cost of its energy storage. However, the optimal policy should take into consideration the future price, and balance the
trade-off between the current operational cost and the future saving in energy purchase. Some works
\cite{Tong_DR_Allerton2012}--\cite{HuangWalrand_QoU_PES} propose the optimal foresighted DSM strategy in the presence of the energy storage under a
\emph{single}-aggregator model. In practice, the power system has many aggregators making decisions that affect each other's cost. In the practical
system with many aggregators, it is inefficient for each aggregator to simply adopt the optimal foresighted strategy designed for a single-aggregator
model. As we will show in the simulation, in the multiple-aggregator scenario, the total cost achieved by such a simple adaptation of the optimal
single-aggregator strategy is much higher (up to 30\%) than the proposed optimal solution. This is because the single-aggregator strategy aims at
achieving individual minimum cost, instead of the total cost. Due to the coupling among the aggregators, the outcome in which individual costs are
minimized may be very different from the outcome in which the total cost is minimized.

In Table~\ref{table:RelatedWork}, we summarize the above discussions on the existing works in power systems by comparing them in various aspects. We
will provide a more technical comparison with the existing works in Table~\ref{table:DetailedComparison_Frameworks}, after we described the proposed
framework.

\begin{table}
\renewcommand{\arraystretch}{1.0}
\caption{Comparisons With Related Mathematical Frameworks.} \label{table:RelatedWork_MathematicalFramework}
\begin{tabular}{|c|c|c|c|c|c|c|}
\hline
\multirow{2}{*}{} & \multirow{2}{*}{MDP} & MU-MDP               & Lyapunov Optimization                                  & Stochastic Control & Stochastic Games             & \multirow{2}{*}{This work} \\
                  &                      &\cite{Hawkins2003}\cite{FuVDS_JSAC2010} & \cite{Tong_DR_Allerton2012}--\cite{HuangWalrand_QoU_PES} & \cite{Altman}      & \cite{HornerSugayaTakahashi} & \\
\hline
Number of decision makers & Single & Multiple & Single & Multiple & Multiple & Multiple \\
\hline
Decentralized information & N/A & Yes & N/A & Yes & No & Yes \\
\hline
Coupling among users & N/A & Weak & N/A & Strong & Strong & Strong \\
\hline
Optimal & Yes & Yes & Yes & No & Yes & Yes \\
\hline
Constructive & Yes & Yes & Yes & Yes & No & Yes \\
\hline
\end{tabular}
\end{table}

\subsection{Related Theoretical Frameworks}
Decision making in a dynamically changing environment has been studied and formulated as Markov decision processes (MDPs). Most MDPs solve for
single-user decision problems. There have been few works \cite{Hawkins2003}\cite{FuVDS_JSAC2010} on multi-user MDPs (MU-MDPs). The works on MU-MDPs
\cite{Hawkins2003}\cite{FuVDS_JSAC2010} focus on \emph{weakly coupled} MU-MDPs, where the term ``weakly coupled'' is coined by \cite{Hawkins2003} to
denote the assumption that one user's action does not directly affect the others' current payoffs. The users are coupled only through some linking
constraints on their actions (for example, the sum of their actions, say the sum data rate, should not exceed some threshold, say the available
bandwidth). However, once a user chooses its own action, its current payoff and its state transition are determined and do not depend on the other
users' actions. In contrast, in this work, the users are \emph{strongly coupled}, namely one user's action directly affect the others' current
payoffs. For example, one aggregator's energy purchase affects the unit price of energy, which has impact on the other aggregators' payments. There
are few works in stochastic control that model the users' interaction as strongly coupled \cite{Altman}. However, the main focus of \cite{Altman} is
to prove the existence of a Nash equilibrium (NE) strategy. There is no performance analysis/guarantee of the proposed NE strategy.

The interaction among users with strong coupling is modeled in the game theory literature as a stochastic game \cite{HornerSugayaTakahashi} in game
theory literature. However, in standard stochastic games, the state of the system is known to all the players. Hence, we cannot model the interaction
of entities in our work as a stochastic game, because different entities have different private states unknown to the others. In addition, the
results in \cite{HornerSugayaTakahashi} are not constructive. They focus on \emph{what} payoff profiles are achievable, but cannot show \emph{how} to
achieve those payoff profiles. In contrast, we propose an algorithm to compute the optimal strategy profile.

In Table~\ref{table:RelatedWork_MathematicalFramework}, we compare our work with existing theoretical frameworks. Note that we will provide a more
technical comparison with the Lyapunov optimization and MU-MDP frameworks in Table~\ref{table:DetailedComparison_Frameworks}, after we described the
proposed framework.

The rest of the paper is organized as follows. We introduce the system model in Section~\ref{sec:model}, and then formulate the design problem in
Section~\ref{sec:Problem}. We describe the proposed optimal decentralized DSM strategy in Section~\ref{sec:DesignFramework}. Through simulations, we
validate our theoretical results and demonstrate the performance gains of the proposed strategy in Section~\ref{sec:Simulation}. Finally, we conclude
the paper in Section~\ref{sec:Conclusion}.

\section{System Model}\label{sec:model}

We first describe the basic model that will be used in most of the paper, and then discuss some extensions of the model.

\subsection{The Basic Model}
\begin{figure}
\centering
\includegraphics[width =3.0in]{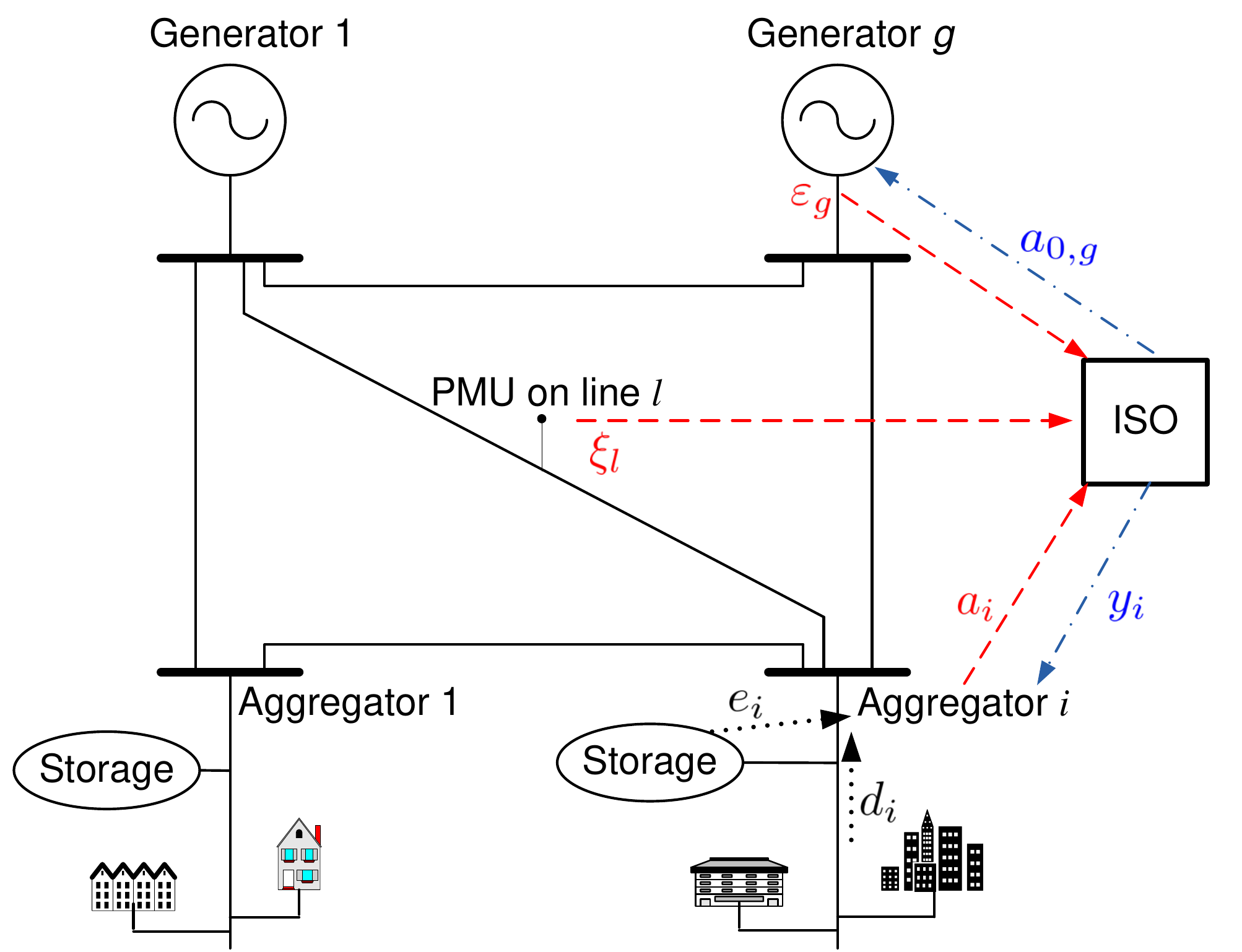}
\caption{The system model of the smart grid. The information flow to the ISO is denoted by red dashed lines, the information flow to the aggregators
is denoted by black dotted lines, and the information flow sent from the ISO is denoted by blue dash-dot lines.} \label{fig:SystemModel}
\end{figure}

We consider a smart grid with one ISO indexed by $0$, $G$ generators indexed by $g=1,2,\ldots,G$, $I$ aggregators indexed by $i=1,2,\ldots,I$, and
$L$ transmission lines (see Fig.~\ref{fig:SystemModel} for an illustration). The ISO schedules the energy generation of generators and determines the
unit prices of energy for the aggregators. The generators provide the ISO with the information of their energy generation cost functions, based on
which the ISO can minimize the total cost of the system. Since the ISO determines how much energy each generator should produce, we do not model
generators as decision makers in the system; instead, we abstract them by their energy generation cost functions. Each aggregator, equipped with
energy storage, manages the electricity usage in a small community of residential households or a commercial building, and determines how much energy
to buy from the ISO. In summary, the decision makers (or the entities) in the system are the ISO and the $I$ aggregators. We denote the set of
aggregators by $\mathcal{I}=\{1,\ldots,I\}$. In the following, we may refer to the ISO or an aggregator generally as entity $i \in
\{0\}\cup\mathcal{I}$, with entity $0$ being the ISO and entity $i \in \mathcal{I}$ being aggregator $i$.

As discussed before, different entities have different sets of local information, which are modeled as their states. Specifically, the ISO receives
reports of the energy generation cost functions, denoted by $\bm{\varepsilon}=(\varepsilon_1,\ldots,\varepsilon_G)$, from the generators, and
measures the status of the transmission lines such as the phases, denoted by $\bm{\xi}=(\xi_1,\ldots,\xi_L)$, by using the phasor measurement units
(PMUs). We summarize the energy generation cost functions and the status of the transmission lines into the ISO's state $s_0 = (\bm{\varepsilon},
\bm{\xi}) \in S_0$, which is unknown to the aggregators\footnote{In some systems, the ISO will provide information about the energy generation cost
and the state of the grid for the aggregators. Our framework still works for this case when the ISO's state $s_0$ is known to the aggregator, as long
as the ISO's state is independent of the aggregators' states.}. Each aggregator receives energy consumption requests from its customers, and manages
its energy storage. We summarize the aggregate demand $d_i$ from aggregator $i$'s customers and the amount $e_i$ of energy in aggregator $i$'s
storage into aggregator $i$'s state $s_i = (d_i, e_i) \in S_i$, which is known to aggregator $i$ only. We assume that all the sets of states are
finite. We highlight which information is available to which entity in Table~\ref{table:InformationAvailability}.

\begin{table}\scriptsize
\renewcommand{\arraystretch}{1.1}
\caption{Information available to each entity.} \label{table:InformationAvailability} \centering
\begin{tabular}{|c|c|}
\hline
Information & Known to whom \\
\hline
$s_0 = (\bm{\varepsilon}, \bm{\xi})$, namely the generation cost functions and the status of the transmission lines & the ISO only \\
\hline
$s_i = (d_i, e_i)$, namely the demand and the amount of energy in storage & Aggregator $i$ only \\
\hline
\end{tabular}
\end{table}

The ISO's action is how much energy each generator should produce, denoted by $a_0 \in A_0 \subset \mathbb{R}_+^G$, where $A_0$ is the action set.
Each aggregator $i$'s action is how much energy to purchase from the ISO, denoted by $a_i \in A_i \subset \mathbb{R}_+$, where $A_i$ is the action
set. We denote the joint action profile of the aggregators as $\bm{a}=(a_1,\ldots,a_N)$, and the joint action profile of all the aggregators other
than $i$ as $\bm{a}_{-i}$.

\begin{figure}
\centering
\includegraphics[width =3.2in]{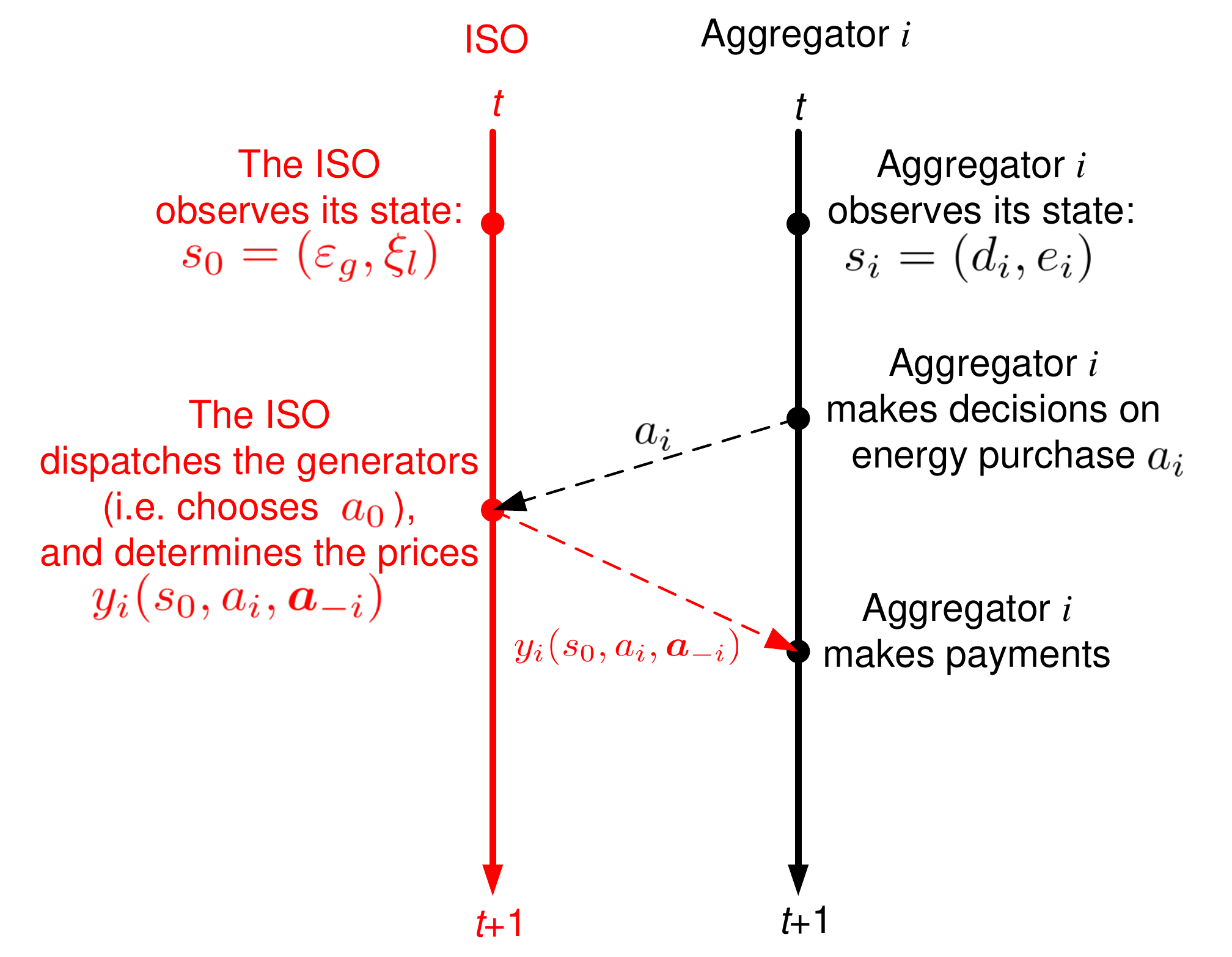}
\caption{Illustration of the interaction between the ISO and aggregator $i$ (i.e. their decision making and information exchange) in one period.}
\label{fig:TimeLine}
\end{figure}

We divide time into periods $t=0,1,2,\ldots$, where the duration of a period is determined by how fast the demand or supply changes or how frequently
the energy trading decisions are made. In each period $t$, the entities act as follows (see Fig.~\ref{fig:TimeLine} for illustration):
\begin{itemize}
\item The ISO observes its state $s_0$.
\item Each aggregator $i$ observes its state $s_i$.
\item Each aggregator $i$ chooses its action $a_i$, namely how much energy to purchase from the ISO, and tells its amount $a_i$ of energy purchase
to the ISO.
\item Based on its state $s_0$ and the aggregators' action profile $\bm{a}$, the ISO determines the price\footnote{We do not model the pricing as the ISO's action, because it does not affect the ISO's payoff, i.e. the social welfare (this is because the payment from the aggregators to the ISO is a monetary transfer within the system and does not count in the social welfare).} $y_i(s_0,\bm{a}) \in Y_i$ of electricity at each aggregator $i$, and announces it to each aggregator
$i$. The ISO also determines its action $a_0$, namely how much energy each generator should produce.
\item Each aggregator $i$ pays $y_i(s_0,\bm{a}) \cdot a_i$ to the ISO.\footnote{Since we consider the interaction among the ISO and the aggregators only, we neglect the payments from the ISO to the generators, which are not included in the total cost anyway, because the payments are transferred among the entities in the system.}
\end{itemize}

The instantaneous cost of each entity depends on its current state and its current action. Each aggregator $i$'s total cost consists of two parts:
the operational cost and the payment. Each aggregator $i$'s operational cost $c_i: S_i \times A_i \rightarrow \mathbb{R}$ is a convex increasing
function of its action $a_i$. An example operational cost function of an aggregator can be
$$
c_i(s_i, a_i) = p \cdot \bm{1}_{\{e_i+a_i < d_i\}} + m_i(e_i),
$$
where $\bm{1}_{\{\cdot\}}$ is the indicator function, $p>0$ is a large positive number that is the penalty when the demand is not fulfilled (i.e.
when $e_i+a_i < d_i$), and $m_i(e_i)$ is the maintenance cost of the energy storage that is convex \cite{ChandyLowTopcuXu}. Then we write each
aggregator $i$'s total cost, which is the cost aggregator $i$ aims to minimize, as the sum of the operational cost and the payment, namely $\bar{c}_i
= c_i + y_i(s_0, a_i, \bm{a}_{-i}) \cdot a_i$, where $\bm{a}_{-i}$ is the actions of the other aggregators. Note that each aggregator's payments
depends on the others' actions through the price. Although each aggregator $i$ observes its realized price $y_i$, it does not know how its action
$a_i$ influences the price $y_i$, because the price depends on the others' actions $\bm{a}_{-i}$ and the ISO's state $s_0$, neither of which is known
to aggregagtor $i$.

The energy generation cost of generator $g$ is denoted $c_g(\varepsilon_g,a_{0,g})$, which is assumed to be convex increasing in the energy
production level $a_{0,g}$. An example cost function can be
$$
c_g(\varepsilon_g,a_{0,g}) = (q_{0,g} + q_{1,g} \cdot a_{0,g} + q_{2,g} \cdot a_{0,g}^2) + q_{r,g} \cdot (a_{0,g}-a_{0,g}^-)^2,
$$
where $a_{0,g}^-$ is the production level in the previous time slot. In this case, the energy generation cost function of generator $g$ is a vector
$\varepsilon_g=(q_{0,g},q_{1,g},q_{2,g},q_{r,g},a_{0,g}^-)$. In the cost function, $q_{0,g} + q_{1,g} \cdot a_{0,g} + q_{2,g} \cdot a_{0,g}^2$ is the
quadratic cost of producing $a_0$ amount of energy \cite{Mohsenian-Rad_TransSmartGrid}\cite{LiChenLow}, and $q_{r,g} \cdot (a_{0,g}-s_{0,g})^2$ is
the ramping cost of changing the energy production level. We denote the total generation cost by $c_0=\sum_{g=1}^G c_g$. The ISO's cost, denoted
$\bar{c}_0$, is then the sum of generation costs and the aggregators' costs, i.e. $\bar{c}_0 = \sum_{i=0}^N c_i$.

We assume that each entity's state transition is Markovian, namely its current state depends only on its previous state and its previous action.
Under the Markovian assumption, we denote the transition probability of entity $i$'s state $s_i$ by $\rho_i(s_i^\prime|s_i,a_i)$. This assumption
holds for the ISO for the following reasons. The ISO's state consists of the energy generation cost functions and the status of the transmission
lines. For renewable energy generation, the energy generation cost function is modeled by the amount of available renewable energy sources (e.g. the
wind speed in wind energy, and the amount of sunshine in solar energy), which is usually assumed to be i.i.d.
\cite{JiangLow}\cite{ScutariPalomar2013a}\cite{ScutariPalomar2013b}. In our model, we relax the i.i.d. assumption and allow the amount of available
renewable energy sources to be correlated across adjacent periods. For conventional energy generation, the energy generation cost function is usually
constant when we do not consider ramping costs. If we consider ramping costs, we can include the energy production level at the previous period in
the energy generation cost function. For the aggregators, the amount of energy left in the storage depends only on the amount of energy in the
previous period and the amount of energy purchases in the current period. The demand of the aggregator is the total demand of all its customers.
Since the number of customers is large, the temporal correlation of each customer's energy demand can be neglected in the total demand. For this
reason, the demand of the aggregator is often assumed to be i.i.d. \cite{HuangWalrand_EnergyStorage_TR}\cite{HuangWalrand_QoU_PES}. In our model, we
relax the i.i.d. assumption and allow the demand of the aggregator to be temporally correlated across adjacent periods.

We also assume that conditioned on the ISO's action $a_0$ and the aggregators' action profile $\bm{a}$, each entity's state transition is independent
of each other. This assumption holds for the ISO, because the energy generation cost functions and the status of the transmission lines depend on the
environments such as weather conditions, and possibly on the previous energy production levels when we consider ramping costs, but not on the
aggregators' demand or its energy storage. For each aggregator, its energy storage level depends only on its own state and action, but not on the
ISO's or the other aggregators' states. The demand of each aggregator could potentially depend on the ISO's state, because the ISO's state influences
the unit price of energy. However, in practice, consumers are not exposed to real-time pricing in most cases, and hence are not price-anticipating
(namely they do not determine how much to consume based on their anticipation of the real-time prices). As a result, it is reasonable to assume that
the demand of each aggregator is independent of the ISO's and the other aggregators' states.

\subsection{Discussions and Extensions}
\subsubsection{Non-Stationarity} An important concern in smart grids is that the demand is non-stationary, namely the demand is significantly higher
in peak hours. In this case, the Markovian assumption on the aggregators' states would not hold if we used the same definition of state. However, we
can augment the state of each aggregator with a component $h$ that represents the period in one day. Then the newly-defined state transition is
Markovian. Similarly, the difference of demand in weekdays and weekends can be captured in the same way, where the additional component $h$ makes the
state space even larger. Note that the seasonal changes in demand cannot be modeled in this way, which would result in a significant increase in the
state space and make the model intractable. However, we can deal with the non-stationarity in such a large time scale by adjusting the system
parameters and recalculating the optimal DSM strategy when the system parameters change.

\subsubsection{Two-Settlement Markets} Many energy markets (such as the New England market and the PJM market) are two-settlement markets.
Specifically, each aggregator predicts the demand in the next day and submits purchase requests in each period of the next day in advance. In this
case, each aggregator takes actions at two time scales: day-ahead and real-time. We can model the two-settlement market by modeling the day-ahead
purchase as a state of the aggregator, if the aggregator predicts the future demand based on historical statistics. Specifically, aggregator $i$'s
state is $s_i=(h,\underline{d}_i,d_i,e_i)$, where $h$ is the period in the day, $\underline{d}_i$ is the amount of energy purchased day-ahead for
period $h$, and $d_i$ is the real-time demand in period $h$. The amount $\underline{d}_i$ of energy purchased day-ahead for period $h$ depends on the
period $h$ of the day. In this case, aggregator $i$ only needs to fulfill the residual demand $d_i-\underline{d}_i$ in real time. This approach to
model the two-settlement market is also adopted in \cite{ZhaoTopcuLow}.

\subsection{The DSM Strategy}
At the beginning of each period $t$, each aggregator $i$ chooses an action based on all the information it has, namely the history of its private
states and the history of its prices. We write each aggregator $i$'s history in period $t$ as $h_i^t=(s_i^0,y_i^0; s_i^1,y_i^1; \ldots;
s_i^{t-1},y_i^{t-1}; s_i^t)$, and the set of all possible histories of aggregator $i$ in period $t$ as $\mathcal{H}_i^t=S_i^{t+1} \times Y_i^t$.
Hence, each aggregator $i$'s strategy can be written as $\pi_i : \cup_{t=0}^\infty \mathcal{H}_i^t \rightarrow A_i$. Similarly, we write the ISO's
history in period $t$ as $h_0^t=(s_0^0,\bm{y}^0; s_0^1,\bm{y}^1; \ldots; s_0^{t-1},\bm{y}^{t-1}; s_0^t)$, where $\bm{y}^t$ is the collection of
prices at period $t$, and the set of all possible histories of the ISO in period $t$ as $\mathcal{H}_0^t=S_0^{t+1} \times \prod_{i\in\mathcal{N}}
Y_i^t$. Then the ISO's strategy can be written as $\pi_0 : \cup_{t=0}^\infty \mathcal{H}_i^t \rightarrow A_i$. The joint strategy profile of all the
entities is written as $\bm{\pi}=(\pi_1,\ldots,\pi_N)$. Since each entity's strategy depends only on its local information, the strategy $\bm{\pi}$
is \emph{decentralized}. Among all the decentralized strategies, we are interested in stationary decentralized strategies, in which the action to
take depends only on the current information, and this dependence does not change with time. Specifically, entity $i$'s \emph{stationary} strategy is
a mapping from its set of states to its set of actions, namely $\pi_i^s : S_i \rightarrow A_i$. Since we focus on stationary strategies, we drop the
superscript $s$, and write $\pi_i$ as entity $i$'s stationary strategy.

The joint strategy profile $\bm{\pi}$ and the initial state $(s_0^0,s_1^0,\ldots,s_N^0)$ induce a probability distribution over the sequences of
states and prices, and hence a probability distribution over the sequences of total costs $\bar{c}_i^0, \bar{c}_i^1,\ldots$. Taking expectation with
respect to the sequences of stage-game payoffs, we have entity $i$'s expected long-term cost given the initial state as
\begin{eqnarray}\label{eqn:LongTermAggregatorCost}
\bar{C}_i(\bm{\pi}|(s_0^0,s_1^0,\ldots,s_I^0)) = \mathbb{E} \left\{ (1-\delta) \sum_{t=0}^{\infty} \left(\delta^t \cdot \bar{c}_i^t\right) \right\},
\end{eqnarray}
where $\delta \in [0,1)$ is the discount factor.

\section{The Design Problem} \label{sec:Problem}
The designer, namely the ISO, aims to maximize the social welfare, namely minimize the long-term total cost in the system. In addition, we need to
satisfy the constraints due to the capacity of the transmission lines, the supply-demand requirements, and so on. We denote the constraints by
$\bm{f}(s_0,a_0,\bm{a})\leq\bm{0}$, where $\bm{f}(s_0,a_0,\bm{a}) \in \mathbb{R}^N$ with $N$ being the number of constraints. We assume that the
electricity flow can be approximated by the direct current (DC) flow model, in which case the constraints $\bm{f}(s_0,a_0,\bm{a})\leq\bm{0}$ are
linear in each $a_i$. Hence, the design problem can be formulated as
\begin{eqnarray}\label{eqn:DesignProblem}
& \min_{\bm{\pi}} & \sum_{s_0^0,s_1^0,\ldots,s_I^0} \left\{C_0(\bm{\pi}|(s_0^0,s_1^0,\ldots,s_I^0)) + \sum_{i\in\mathcal{I}}
C_i(\bm{\pi}|(s_0^0,s_1^0,\ldots,s_I^0))\right\} \\
& s.t. & \bm{f}(s_0,\pi_0(s_0),\pi_1(s_1),\ldots,\pi_I(s_I))\leq\bm{0}, ~\forall (s_0,s_1,\ldots,s_N). \nonumber
\end{eqnarray}
Note that in the above optimization problem, we use aggregator $i$'s cost $C_i$ instead of its total cost $\bar{C}_i$, because its payment is
transferred to the ISO and is thus canceled in the total cost. Note also that we sum up the social welfare under all the initial states. This can be
considered as the expected social welfare when the initial state is uniformly distributed. The optimal stationary strategy profile that maximizes
this expected social welfare will also maximize the social welfare given any initial state. We write the solution to the design problem as
$\bm{\pi}^\star$ and the optimal value of the design problem as $C^\star$.

\section{Optimal Foresighted Demand Side Management} \label{sec:DesignFramework}
In this section, we derive the optimal foresighted DSM strategy assuming that each entity knows its own state transition probabilities.

\subsection{The aggregator's Decision Problem and Its Conjectured Price}
Contrary to the designer, each aggregator aims to minimize its own long-term total cost $\bar{C}_i(\bm{\pi}|(s_0^0,s_1^0,\ldots,s_N^0))$. In other
words, each aggregator $i$ solves the following problem:
$$
\pi_i = \arg\max_{\pi_i^\prime} \bar{C}_i(\pi_i^\prime,\bm{\pi}_{-i}|(s_0^0,s_1^0,\ldots,s_N^0)).
$$
Assuming that the aggregator knows all the information, the optimal solution to the above problem should satisfy the following:
$$
V(s_0,s_i,\bm{s}_{-i}) = \max_{a_i \in A_i} (1-\delta) \bar{c}_i(s_0,s_i,a_i,\bm{a}_{-i}) + \delta \cdot
\sum_{s_0^\prime,s_i^\prime,\bm{s}_{-i}^\prime} \left\{\rho_0(s_0^\prime|s_0) \prod_{j \in \mathcal{N}} \rho_j(s_j^\prime|s_j,a_j)
V(s_0^\prime,s_i^\prime,\bm{s}_{-i}^\prime)\right\}.
$$
Note that the above equations would be the Bellman equations, if the aggregator knew all the information such as the other aggregators' strategies
$\bm{\pi}_{-i}$ and states $\bm{s}_{-i}$, and the ISO's state $s_0$. However, such information is never known to the aggregator. Hence, we need to
separate the influence of the other entities from each aggregator's decision problem.

One way to decouple the interaction among the aggregators is to endow each aggregator with a conjectured price. In general, the conjecture informs
the aggregator of what price it should anticipate given its state and its action. However, in the presence of decentralized information, such a
complicated conjecture is hard, if not possible, to form. Specifically, aggregator $i$'s conjectured price should depend not only on aggregator $i$'s
action and state, but also on the ISO's state. Hence, no entity possess \emph{all the necessary information} to form the conjecture. For this reason,
in this paper, we propose a simple conjecture, namely the price does not depend on the aggregator's state and action. In this case, the conjectures
can be formed by the ISO based on its local information and then communicated to the aggregators. Denote the conjectured price as $\tilde{y}_i$, we
can rewrite aggregator $i$'s decision problem as
$$
\tilde{V}^{\tilde{y}_i}(s_i) = \max_{a_i \in A_i} (1-\delta) \left[ c_i(s_i,a_i) + \tilde{y}_i \cdot a_i \right] + \delta \cdot \sum_{s_i^\prime}
\left[\rho_i(s_i^\prime|s_i,a_i) \tilde{V}^{\tilde{y}_i}(s_i^\prime)\right].
$$
Clearly, we can see from the above equations that given the conjectured price $\tilde{y}_i$, each aggregator can make decisions based only on its
local information.

\begin{figure}
\centering
\includegraphics[width =3.2in]{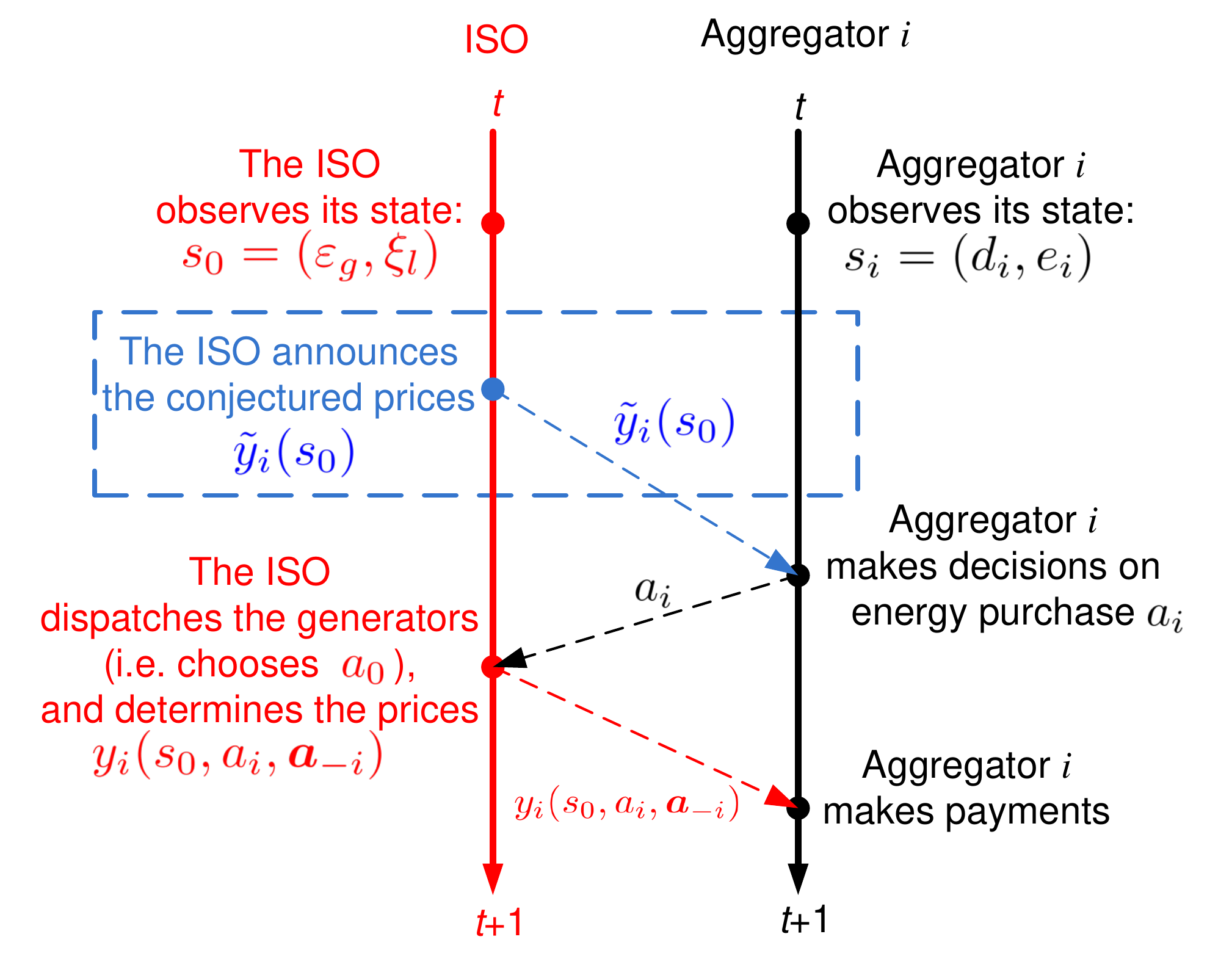}
\caption{Illustration of the entities' decision making and information exchange in the design framework based on conjectured prices.}
\label{fig:TimeLine_Conjecture}
\end{figure}

In Fig.~\ref{fig:TimeLine_Conjecture}, we illustrate the entities' decision making and information exchange in the design framework based on
conjectured prices. Comparing Fig.~\ref{fig:TimeLine_Conjecture} with Fig.~\ref{fig:TimeLine} of the system without conjectured prices, we can see
that in the proposed design framework, the ISO sends the conjectured prices to the aggregators before the aggregators make decisions. This additional
procedure of exchanging conjectured prices allows the ISO to lead the aggregators to the optimal DSM strategies. Note that the conjectured price is
generally not equal to the real price charged at the end of the period, and is not equal to the expectation of the real price in the future. In this
sense, the conjectured prices can be considered as control signals sent from the ISO to the aggregators, which can help the aggregators to compute
the optimal strategies. In Section~\ref{sec:Simulation}, we will compare the conjectured price with the expected real price by simulation.

The remaining question is how to determine the optimal conjectured prices, such that when each aggregator reacts based on its conjectured price, the
resulting strategy profile maximizes the social welfare.

\subsection{The Optimal Decentralized DSM Strategy}
The optimal conjectured prices depend on the ISO's state, which is known to the ISO only. Hence, we propose a distributed algorithm used by the ISO
to iteratively update the conjectured prices and by the aggregators to update their optimal strategies. The algorithm will converge to the optimal
conjectured prices and the optimal strategy profile that achieves the minimum total system cost $C^\star$.

At each iteration $k$, given the conjectured price $\tilde{y}_i^{(k)}$, each aggregator $i$ solves
$$
\tilde{V}_i^{\tilde{y}_i^{(k)}}(s_i) = \max_{a_i \in A_i} (1-\delta) \left[ c_i(s_i,a_i) + \tilde{y}_i^{(k)} \cdot a_i \right] + \delta \cdot
\sum_{s_i^\prime} \left[\rho_i(s_i^\prime|s_i,a_i) \tilde{V}_i^{\tilde{y}_i^{(k)}}(s_i^\prime)\right],
$$
and obtains the optimal value function $\tilde{V}_i^{\tilde{y}_i^{(k)}}$ as well as the corresponding optimal strategy $\pi_i^{\tilde{y}_i^{(k)}}$
under the current conjectured price $\tilde{y}_i^{(k)}$.

Similarly, given the conjectured prices $\bm{\tilde{y}}_0^{(k)} \in \mathbb{R}^G$, the ISO solves
$$
\tilde{V}_0^{\bm{\tilde{y}}_0^{(k)}}(s_0) = \min_{a_i \in A_i} (1-\delta) \left[\sum_g c_g(s_0,a_0) + \bm{\tilde{y}}_0^{(k)T} \cdot a_0
\right]+\delta \cdot \sum_{s_0^\prime} \left[\rho_0(s_0^\prime|s_0,a_0) \tilde{V}_0^{\bm{\tilde{y}}_0^{(k)}}(s_0^\prime)\right],
$$
and obtains the optimal value function $\tilde{V}_0^{\tilde{y}_i^{(k)}}$ as well as the corresponding optimal strategy $\pi_0^{\tilde{y}_i^{(k)}}$
under the current conjectured price $\tilde{y}_i^{(k)}$.

Then the ISO updates the conjectured prices as follows:
$$
\tilde{y}_i^{(k+1)} = \left(\bm{\lambda}^{(k+1)}(s_0)\right)^T \cdot \frac{\partial f(s_0,\bm{a})}{\partial a_i},
$$
where $\bm{\lambda}^{(k+1)}(s_0)\in\mathbb{R}^N$ is calculated as
$$
\bm{\lambda}^{(k+1)}(s_0) = \left\{\bm{\lambda}^{(k)}(s_0) + \Delta(k) \cdot \bm{f}\left(s_0, \pi_0^{\bm{\tilde{y}}_0(k)}(s_0),\frac{\sum_{s_1}
\pi_1^{\tilde{y}_1(k)}(s_1)}{|S_1|},\ldots,\frac{\sum_{s_I} \pi_I^{\tilde{y}_I(k)}(s_I)}{|S_I|}\right) \right\}^+,
$$
where $\Delta(k) \in \mathbb{R}_{++}$ is the step size, and $\{x\}^+=\max\{x,0\}$.

Note that in the above update of conjectures, to calculate (the subgradient) $\bm{\lambda}^{(k)}$, the ISO needs to know the average amount of
purchase $\frac{\sum_{s_i} \pi_i^{\tilde{y}_1(k)}(s_i)}{|S_i|}$ from each aggregator $i$. This requires additional information exchange from the
aggregator to the ISO. Moreover, the aggregator may not be willing to report such information to the ISO. To reduce the amount of information
exchange and preserve privacy, we propose that the ISO calculates the empirical mean values of the aggregators' purchases in the run-time (which
results in stochastic subgradients). We summarize the algorithm in Table~\ref{table:DecentralizedStrategy}, and prove that the algorithm can achieve
the optimal social welfare in the following theorem.

\begin{theorem}\label{theorem:Convergence}
The algorithm in Table~\ref{table:DecentralizedStrategy} converges to the optimal strategy profile, namely
$$
\lim_{k\rightarrow0} \left| \sum_{s_0^0,s_1^0,\ldots,s_I^0} \left\{C_0(\bm{\pi}^{\tilde{y}^{(k)}}|(s_0^0,s_1^0,\ldots,s_I^0)) +
\sum_{i\in\mathcal{I}} C_i(\bm{\pi}^{\tilde{y}^{(k)}}|(s_0^0,s_1^0,\ldots,s_I^0))\right\} - C^\star \right| = 0.
$$
\end{theorem}
\begin{IEEEproof}
See the appendix.
\end{IEEEproof}

\begin{table}
\renewcommand{\arraystretch}{1.3}
\caption{Distributed algorithm to compute the optimal decentralized DSM strategy.} \label{table:DecentralizedStrategy} \centering
\begin{tabular}{l}
\hline
\textbf{Input:} Each entity's performance loss tolerance $\epsilon_i$ \\
\hline
\textbf{Initialization:} Set $k=0$, $\bar{a}_i(0)=0,\forall i\in\mathcal{I}$, $\tilde{y}_i(0)=0,\forall i\in\mathcal{I}\cup\{0\}$. \\
\textbf{repeat} \\
~~~~Each aggregator $i$ solves \\
~~~~~~~~$\tilde{V}_i^{\tilde{y}_i^{(k)}(s_0)}(s_i) = \max_{a_i \in A_i} (1-\delta) \left[ c_i(s_i,a_i) + \tilde{y}_i^{(k)}(s_0) \cdot a_i \right] +
\delta \cdot
\sum_{s_i^\prime} \left[\rho_i(s_i^\prime|s_i,a_i) \tilde{V}_i^{\tilde{y}_i^{(k)}(s_0)}(s_i^\prime)\right]$ \\
~~~~The ISO solves \\
~~~~~~~~$\tilde{V}_0^{\bm{\tilde{y}}_0(k)}(s_0) = \min_{a_i \in A_i} (1-\delta) \left[\sum_g c_g(s_0,a_0) + \bm{\tilde{y}}_0(k)^T \cdot a_0
\right]+\delta \cdot \sum_{s_0^\prime} \left[\rho_0(s_0^\prime|s_0,a_0) \tilde{V}_0^{\bm{\tilde{y}}_0(k)}(s_0^\prime)\right]$ \\
~~~~Each aggregator $i$ reports its purchase request $\pi_i^{\tilde{y}_i^{(k)}(s_0)}(s_i)$ \\
~~~~The ISO updates $\bar{a}_i(k+1)=\bar{a}_i(k) + \pi_i^{\tilde{y}_i^{(k)}(s_0)}(s_i)$ for all $i\in\mathcal{I}$ \\
~~~~The ISO updates the conjectured prices: \\
~~~~~~~~$\tilde{y}_i^{(k+1)}(s_0) = \left(\bm{\lambda}(k+1)^T \cdot \frac{\partial f(s_0,\bm{a})}{\partial a_i}\right)^T$, where $\Delta(k)=\frac{1}{k+1}$ and \\
~~~~~~~~~~~~$ \bm{\lambda}(k+1) = \left\{\bm{\lambda}(k) + \Delta(k) \cdot \bm{f}\left(s_0,
\pi_0^{\bm{\tilde{y}}_0(k)}(s_0),\frac{\bar{a}_1(k+1)}{k+1},\ldots,\frac{\bar{a}_I(k+1)}{k+1}\right) \right\}^+$ \\
\textbf{until} $\|\tilde{V}_i^{\tilde{y}_i^{(k+1)}(s_0)}-\tilde{V}_i^{\tilde{y}_i^{(k)}(s_0)}\| \leq \epsilon_i$ \\
\hline
\end{tabular}
\end{table}

We summarize the information needed by each entity in Table~\ref{table:Information}. We can see that the amount of information exchange at each
iteration is small ($O(I)$), compared to the amount of information unavailable to each entity ($\prod_{j\neq i} |S_i|$ states plus the strategies
$\bm{\pi}_{-i}$). In other words, the algorithm enables the entities to exchange a small amount ($O(I)$) of information and reach the optimal DSM
strategy that achieves the same performance as when each entity knows the complete information about the system.
\begin{table}\scriptsize
\renewcommand{\arraystretch}{1.1}
\caption{Information needed by each entity to implement the algorithm.} \label{table:Information} \centering
\begin{tabular}{|c|c|}
\hline
Entity & Information at each step $k$\\
\hline
The ISO & The purchase request ${\pi_i^{\tilde{y}_i^{(k)}(s_0)}(s_i)}$ of each aggregator \\
\hline
Each aggregator $i$ & Conjecture on its price $\tilde{y}_i^{(k)}(s_0)$ \\
\hline
\end{tabular}
\end{table}

We briefly discuss the complexity of implementing the algorithm in terms of the dimensionality of the Bellman equations solved by each entity. For
each aggregator, it solves the Bellman equation that has the the same dimensionality as the cardinality of its state space, namely $|S_i|$. For each
ISO, the dimensionality of its state space is large, because the generation cost functions $\bm{\varepsilon}$ are a vector of length $G$ and the
status of the transmission lines is a vector of length $L$. However, the ISO's decision problem can be decomposed due to the following observation.
Note that the generators' energy generation cost functions are independent of each other. Then we have the following theorem.

\begin{theorem}
Given the conjectured price $\bm{\tilde{y}}_0(k)$, the ISO's value function $\tilde{V}_0^{\bm{\tilde{y}}_0(k)}$ can be calculated by
$\tilde{V}_0^{\bm{\tilde{y}}_0(k)}(s_0) = \sum_{g=1}^G \tilde{V}_{0,g}^{\tilde{y}_{0,g}(k)}(\varepsilon_{g})$, where
$\tilde{V}_{0,g}^{\tilde{y}_{0,g}(k)}$ solves
$$
\tilde{V}_{0,g}^{\tilde{y}_{0,g}(k)}(\varepsilon_{g}) = \min_{a_{0,g}} (1-\delta) \left[ c_g(\varepsilon_{g},a_{0,g}) + \tilde{y}_{0,g}(k)^T \cdot
a_{0,g} \right]+\delta \cdot \sum_{\varepsilon_{g}^\prime} \left[\rho_0(\varepsilon_{g}^\prime|\varepsilon_{g},a_{0,g})
\tilde{V}_{0,g}^{\tilde{y}_{0,g}(k)}(\varepsilon_{g}^\prime)\right].
$$
\end{theorem}
\begin{IEEEproof}
The proof follows directly from Lemma~\ref{lemma:DecompositionValueFunction} in the appendix.
\end{IEEEproof}

From the above proposition, we know that the dimensionality of the ISO's decision problem is $\sum_{g=1}^G |\mathcal{E}_g|$, where $|\mathcal{E}_g|$
is the cardinality of the set of generator $g$'s generation cost functions. The dimensionality increases linearly with the number of generators,
instead of exponentially with the number of generators and transmission lines without decomposition.


\subsection{Learning Unknown Dynamics}
In practice, each entity may not know the dynamics of its own states (i.e., its own state transition probabilities) or even the set of its own
states. When the state dynamics are not known a priori, each entity cannot solve their decision problems using the distributed algorithm in
Table~\ref{table:DecentralizedStrategy}. In this case, we can adapt the online learning algorithm based on post-decision state (PDS) in
\cite{FuVDS_TVT2009}, which was originally proposed for wireless video transmissions, to our case.

The main idea of the PDS-based online learning is to learn the post-decision value function, instead of the normal value function. Each aggregator
$i$'s post-decision value function is defined as $U_i(\tilde{d}_i,\tilde{e}_i)$, where $(\tilde{d}_i,\tilde{e}_i)$ is the post-decision state. The
difference from the normal state is that the PDS $(\tilde{d}_i,\tilde{e}_i)$ describes the status of the system after the purchase action is made but
before the demand in the next period arrives. Hence, the relationship between the PDS and the normal state is
$$
\tilde{d}_i=d_i,~\tilde{e}_i = e_i + (a_i-d_i).
$$
Then the post-decision value function can be expressed in terms of the normal value function as follows:
\begin{eqnarray*}
U_i(\tilde{d}_i,\tilde{e}_i) = \sum_{d_i^\prime} \rho_i(d_i^\prime,\tilde{e}_i-(a_i-\tilde{d}_i)|\tilde{d}_i,\tilde{e}_i)\cdot V_i(d_i^\prime,
\tilde{e}_i-(a_i-\tilde{d}_i)).
\end{eqnarray*}
In PDS-based online learning, the normal value function and the post-decision value function are updated in the following way:
\begin{eqnarray*}
V_i^{(k+1)}(d_i^{(k)}, e_i^{(k)}) &=& \max_{a_i} (1-\delta) \cdot c_i(d_i^{(k)},e_i^{(k)},a_i) + \delta \cdot U_i^{(k)}(d_i^{(k)},
e_i^{(k)}+(a_i-d_i^{(k)})), \\
U_i^{(k+1)}(d_i^{(k)}, e_i^{(k)}) &=& (1-\alpha^{(k)}) U_i^{(k)}(d_i^{(k)}, e_i^{(k)}) + \alpha^{(k)} \cdot V_i^{(k)}(d_i^{(k)},
e_i^{(k)}-(a_i-d_i^{(k)})).
\end{eqnarray*}
We can see that the above updates do not require any knowledge about the state dynamics. It is proved in \cite{FuVDS_TVT2009} that the PDS-based
online learning will converge to the optimal value function.

\subsection{Detailed Comparisons with Existing Frameworks}
Since we have introduced our proposed framework, we can provide a detailed comparison with the existing theoretical framework. The comparison is
summarized in Table~\ref{table:DetailedComparison_Frameworks}.

First, the proposed framework reduces to the myopic optimization framework when we set the discount factor $\delta=0$. In this case, the problem
reduces to the classic economic dispatch problem.

Second, the Lyapunov optimization framework is closely related to the PDS-based online learning. In fact, it could be considered as a special case of
the PDS-based online learning when we set the post-decision value function as $U_i(s_i)=c_i(s_i,a_i)+(e_i+a_i)^2-e_i^2$, and choose the action that
minimizes the post-decision value function in the run-time. However, the Lyapunov drift in the above post-decision value function depends only on the
status of the energy storage, but not on the demand. In contrast, in our PDS-based online learning, we explicitly considers the impact of the demand
when updating the normal and post-decision value functions.

Finally, the key difference between our proposed framework and the framework for MU-MDP \cite{Hawkins2003}\cite{FuVDS_JSAC2010} is how we penalize
the constraints $f(s_0,a_0,\bm{a})$. In particular, the framework in \cite{Hawkins2003}\cite{FuVDS_JSAC2010}, if directly applied in our model, would
define only one Lagrangian multiplier for all the constraints under different states $s_0$. This leads to performance loss in general
\cite{FuVDS_JSAC2010}. In contrast, we define different Lagrangian multipliers to penalize the constraints under different states $s_0$, and
potentially enable the proposed framework to achieve the optimality (which is indeed the case as have been proved in
Theorem~\ref{theorem:Convergence}).

\begin{table}
\renewcommand{\arraystretch}{1.1}
\caption{Relationship between the proposed and existing theoretical frameworks.} \label{table:DetailedComparison_Frameworks} \centering
\begin{tabular}{|c|c|c|}
\hline
Framework & Relationship & Representative works \\
\hline
Myopic & $\delta=0$ & \cite{JiangLow} \\
\hline
Lyapunov optimization & Aggregator $i$'s post-decision value function $U_i(s_i)=c_i(s_i,a_i)+(e_i+a_i-d_i)^2-e_i^2$ & \cite{HuangWalrand_EnergyStorage_TR}\cite{HuangWalrand_QoU_PES} \\
\hline
MU-MDP & Lagrangian multiplier $\bm{\lambda}(s_0)=\bm{\lambda}$ for all $s_0$ & \cite{Hawkins2003}\cite{FuVDS_JSAC2010} \\
\hline
\end{tabular}
\end{table}

\section{Simulation Results} \label{sec:Simulation}
In this section, we validate our theoretical results and compare against existing DSM strategies through extensive simulations. We use the
widely-used IEEE test power systems with the data (e.g. the topology, the admittances and capacity limits of transmission lines) provided by
University of Washington Power System Test Case Archive \cite{WashingtonTest}. We describe the other system parameters as follows (these system
parameters are used by default; any changes in certain scenarios will be specified):
\begin{itemize}
\item One period is one hour. The discount factor is $\delta=0.99$.
\item The demand of aggregator $i$ at period $t$ is uniformly distributed among the interval
$[d_i(t\mod 24) - \Delta d_i(t\mod 24), d_i(t \mod 24) + \Delta d_i(t \mod 24)]$. In other words, the distribution of demand is time-varying across a
day. We let the peak hours for all the aggregators to be from 17:00 to 22:00. The mean value $d_i(t\mod 24)$ and the range $\Delta d_i(t\mod 24)$ of
aggregator $i$'s demand are described as follows (values are adapted from \cite{FisherONeillFerris}):
\begin{eqnarray}\label{eqn:Simulation_DemandMeanValue}
d_i(t\mod 24)=\left\{\begin{array}{ll} 50+(i-1)\cdot 0.5 ~\mathrm{MW} & \mathrm{if}~t\mod 24 \in [17, 22] \\ 25+(i-1)\cdot 0.5 ~\mathrm{MW} &
\mathrm{otherwise}
\end{array} \right.
\end{eqnarray}
and
\begin{eqnarray}\label{eqn:Simulation_DemandDeviationFromMean}
\Delta d_i(t\mod 24)=\left\{\begin{array}{ll} 5 ~\mathrm{MW} & \mathrm{if}~t\mod 24 \in [17, 22] \\ 2 ~\mathrm{MW} & \mathrm{otherwise}
\end{array} \right.
\end{eqnarray}
\item All the aggregators have energy storage of the same capacity $25$~MW.
\item All the aggregators have the same linear energy storage cost function \cite{ChandyLowTopcuXu}:
\begin{eqnarray*}
c_i(s_i,a_i) = 2 \cdot (a_i-d_i)^+,
\end{eqnarray*}
namely the maintenance cost grows linearly with the remaining energy level and is independent of the amount of charge and discharge.
\item We index the energy generators starting from the renewable energy generators. All the renewable energy generators have linear energy generation cost functions: \cite{FisherONeillFerris}
\begin{eqnarray*}
c_g(a_{0,g}) = g \cdot a_{0,g},
\end{eqnarray*}
where the unit energy generation cost has the same value as the index of the generator (these values are adapted from \cite{FisherONeillFerris},
which cited that the unit energy generation cost ranges from \$0.19/MWh to \$10/MWh). Although the energy generation cost function is deterministic,
the maximum amount of energy production is stochastic (due to wind speed, the amount of sunshine, and so on). The maximum amounts of energy
production of all the renewable energy generators follow the same uniform distribution in the range of $[90, 110]$~MW.
\item The rest of energy generators are conventional energy generators that use coal, all of which have the same energy generation cost function: \cite{ChandyLowTopcuXu}
\begin{eqnarray*}
c_g(a_{0,g}) = \underbrace{\frac{1}{2} (a_{0,g})^2}_{\mathrm{generation~cost}} + \underbrace{\frac{1}{10}
(a_{0,g}-a_{0,g}^-)^2}_{\mathrm{ramping~cost}}.
\end{eqnarray*}
In other words, the conventional energy generators have fixed (i.e. not stochastic) generation cost functions.
\item The status of the transmission lines is their capacity limits. The nominal values of the capacity limits are the same as specified in the data
provided by \cite{WashingtonTest}. In each period, we randomly select a line with equal probability, and decrease its capacity limit by 10\%.
\end{itemize}

We compare the proposed DSM strategies with the following schemes.
\begin{itemize}
\item Centralized optimal strategies (``Centralized''): We assume that there is a central controller who knows everything about the system and solves the long-term cost
minimization problem as a single-user MDP. This scheme serves as the benchmark optimum.
\item Myopic strategies (``Myopic'') \cite{Mohsenian-Rad_TransSmartGrid}--\cite{ScutariPalomar2013b}: In each period $t$, the aggregators myopically minimizes their current costs, and based on their actions, the ISO minimizes the current total generation cost.
\item Single-user Lyapunov optimization (``Lyapunov'') \cite{Tong_DR_Allerton2012}--\cite{HuangWalrand_QoU_PES}: We let each aggregator adopt the stochastic
optimization technique proposed in \cite{Tong_DR_Allerton2012}--\cite{HuangWalrand_QoU_PES}. Based on the aggregators' purchases, the ISO minimizes
the current total generation cost.
\end{itemize}


\subsection{Performance Evaluation}
Now we evaluate the performance of the proposed DSM strategy in various scenarios.

\subsubsection{Impact of the energy storage}
First, we study the impact of the energy storage on the performance of different schemes. We assume that all the generators are conventional energy
generators using fossil fuel, in order to rule out the impact of the uncertainty in renewable energy generation (which will be examined next). The
performance criterion is the total cost per hour normalized by the number of buses in the system. We compare the normalized total cost achieved by
different schemes when the capacity of the energy storage increases from 5~MW to 45~MW.

\begin{figure}
\centering
\includegraphics[width =2.4in]{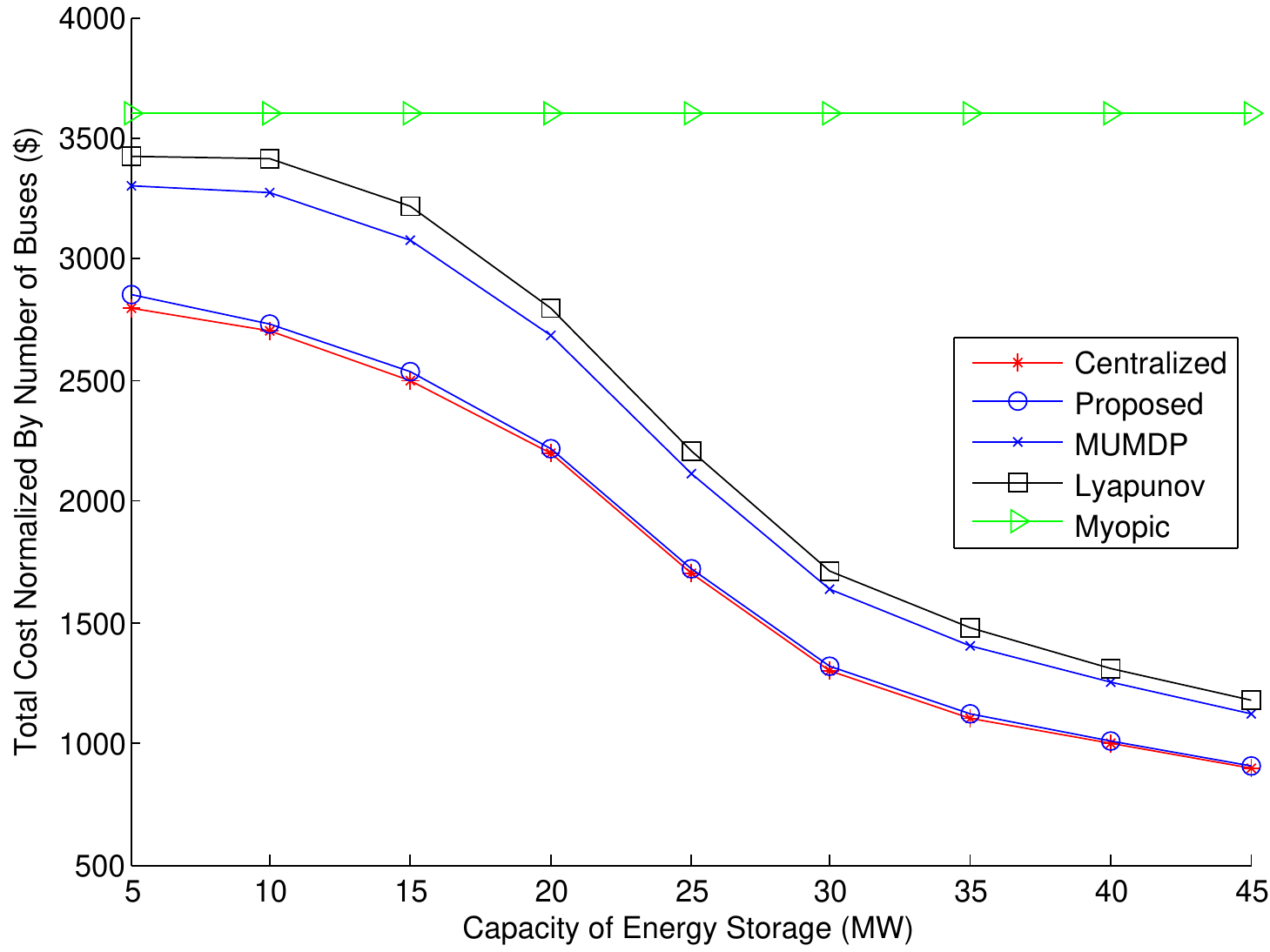}
\caption{The normalized total cost per hour versus the capacity of the energy storage in the IEEE 14-bus system.}
\label{fig:Cost_StorageCapacity_14bus}
\end{figure}

\begin{figure}
\centering
\includegraphics[width =2.4in]{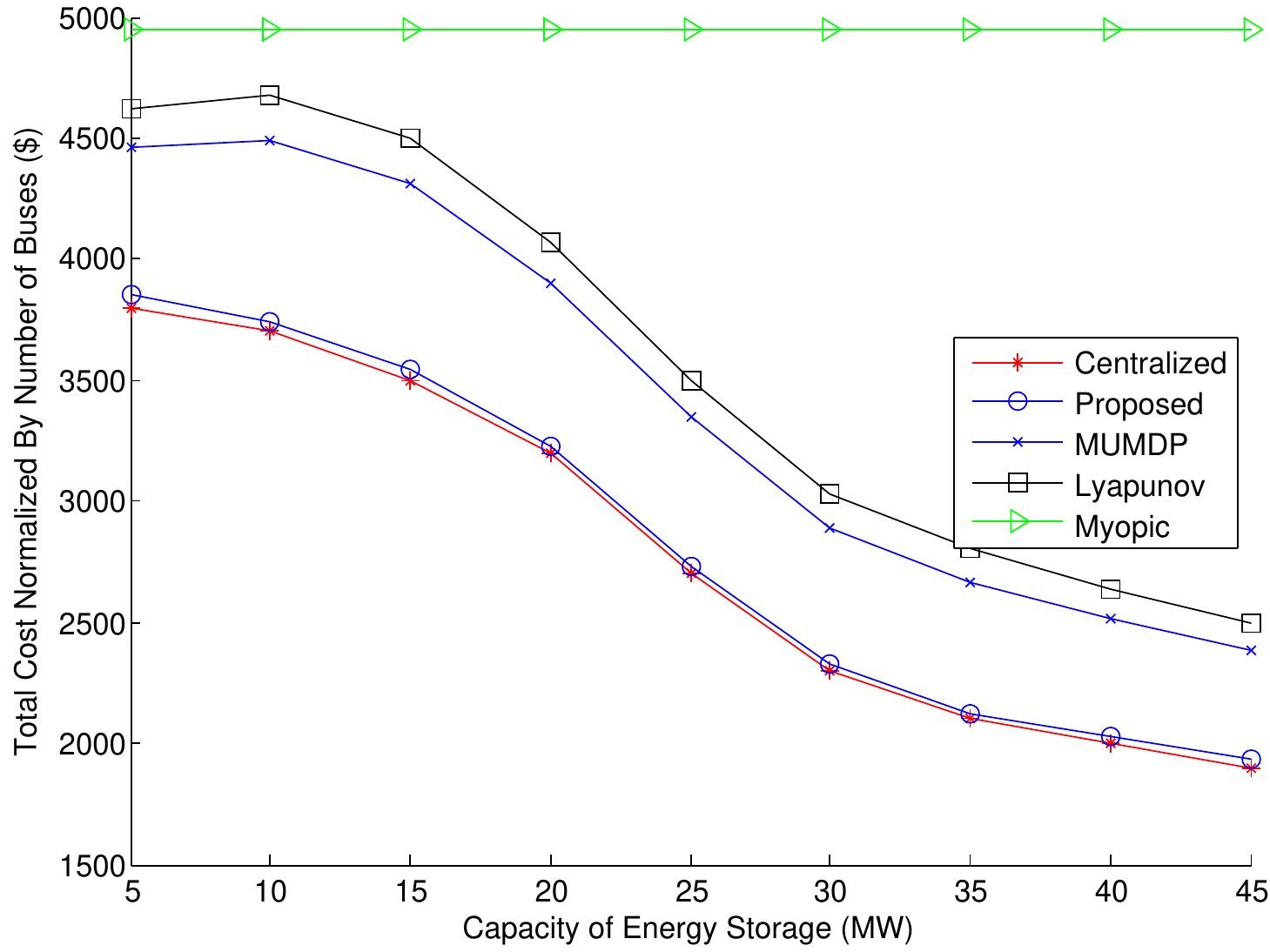}
\caption{The normalized total cost per hour versus the capacity of the energy storage in the IEEE 30-bus system.}
\label{fig:Cost_StorageCapacity_30bus}
\end{figure}

\begin{figure}
\centering
\includegraphics[width =2.4in]{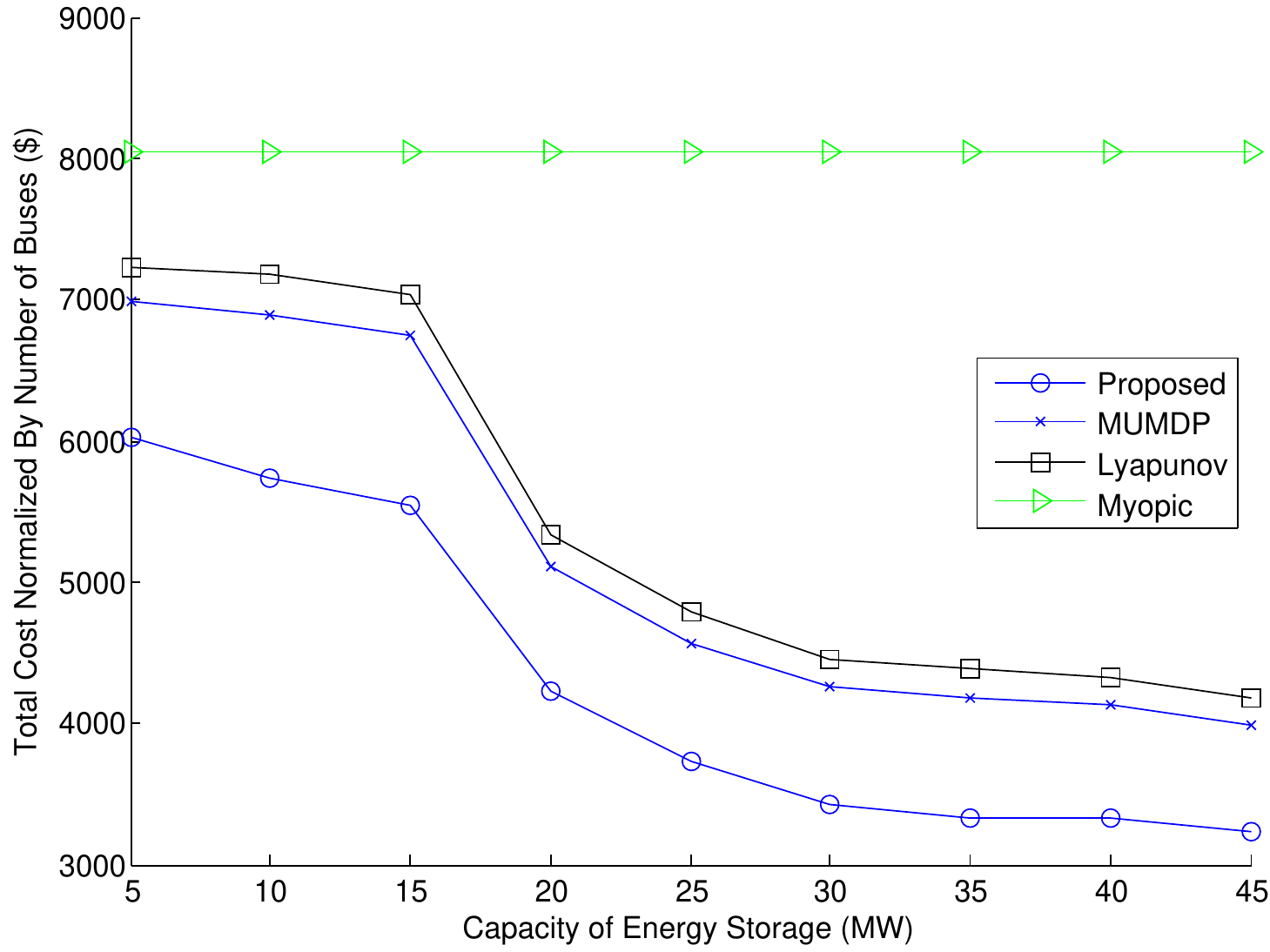}
\caption{The normalized total cost per hour versus the capacity of the energy storage in the IEEE 118-bus system.}
\label{fig:Cost_StorageCapacity_118bus}
\end{figure}

Fig.~\ref{fig:Cost_StorageCapacity_14bus}--\ref{fig:Cost_StorageCapacity_118bus} show the normalized total cost achieved by different schemes under
IEEE 14-bus system, IEEE 30-bus system, and IEEE 118-bus system, respectively. Note that we do not show the performance of the centralized optimal
strategy under IEEE 118-bus system, because the number of states in the centralized MDP is so large that it is intractable to compute the optimal
solution. This also shows the computational tractability and the scalability of the proposed distributed algorithm. Under IEEE 14-bus and 30-bus
systems, we can see that the proposed DSM strategy achieves almost the same performance as the centralized optimal strategy. The slight optimality
gap comes from the performance loss experienced during the convergence process of the conjectured prices. Compared to the DSM strategy based on
single-user Lyapunov optimization, our proposed strategy can reduce the total cost by around 30\% in most cases. Compared to the myopic DSM strategy,
our reduction in the total cost is even larger and increases with the capacity of the energy storage (up to 60\%).

\subsubsection{Impact of the uncertainty in renewable energy generation}
Now we examine the impact of the uncertainty in renewable energy generation. For a given test system, we let half of the generators to be renewable
energy generators. Recall that the maximum amounts of energy production of the renewable energy generators are stochastic and follow the same uniform
distribution. We set the mean value of the maximum amount of energy production to be 100~MW, and vary the range of the uniform distribution. A wider
range indicates a higher uncertainty in renewable energy production. Hence, we define the uncertainty in renewable energy generation as the maximum
deviation from the mean value in the uniform distribution.

\begin{figure}
\centering
\includegraphics[width =3.5in]{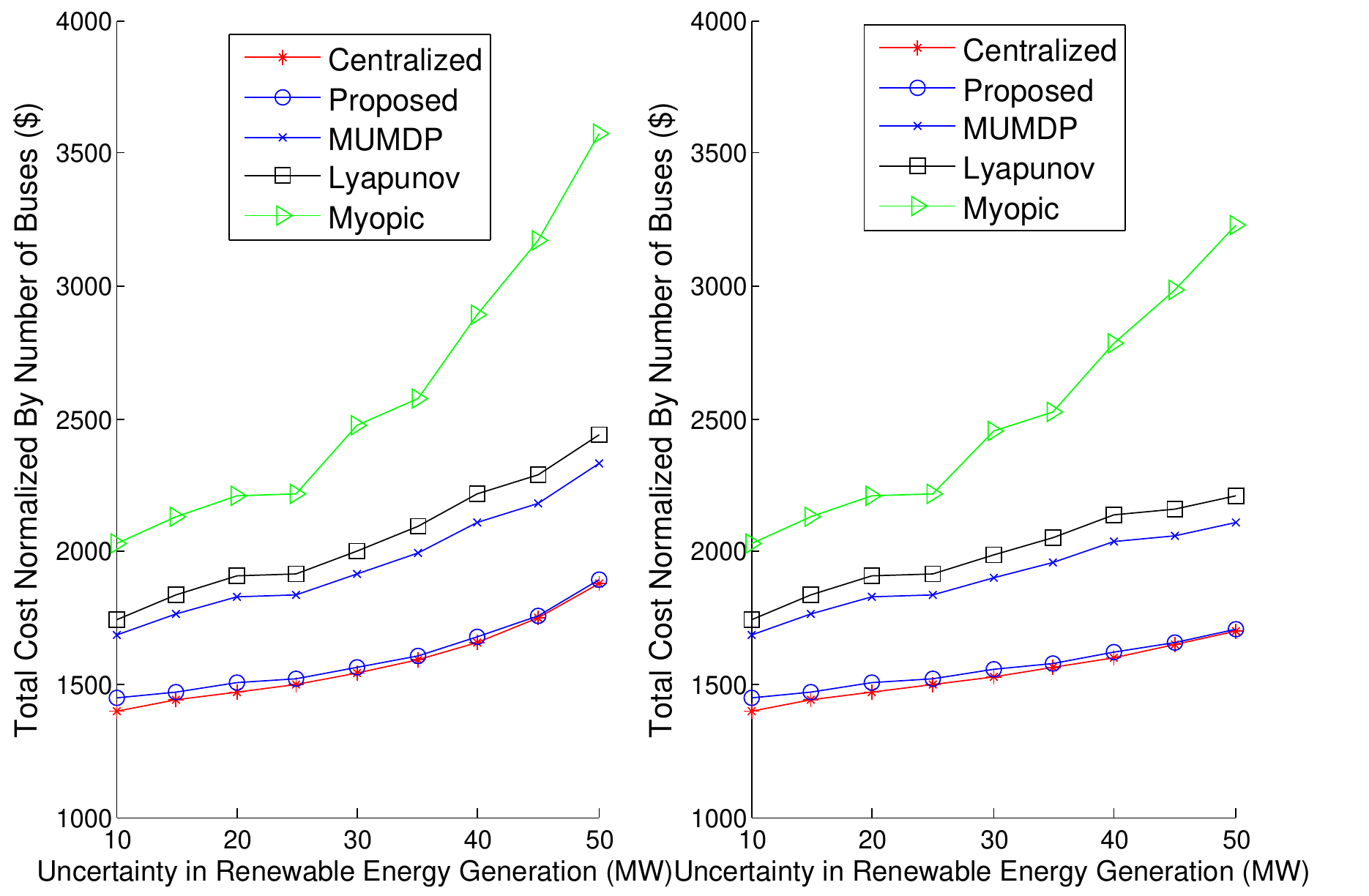}
\caption{The normalized total cost per hour versus the uncertainty in renewable energy generation in the IEEE 14-bus system. The aggregators have
energy storage of capacity 25~MW and 50~MW, respectively.} \label{fig:Cost_Uncertainty_14bus_Storage25_Storage50}
\end{figure}

Fig.~\ref{fig:Cost_Uncertainty_14bus_Storage25_Storage50} shows the normalized total cost under different degrees of uncertainty in renewable energy
generation. Again, the proposed strategy achieves the performance of the centralized optimal strategy in the IEEE 14-bus system. We can see that the
costs achieved by all the schemes increase with the uncertainty in renewable energy generation. This happens for the following reasons. Since the
renewable energy is cheaper, the ISO will dispatch renewable energy whenever possible, and dispatch conventional energy for the residual demand.
However, when the renewable energy generation has larger uncertainty, the variation in the residual demand is higher, which results in a higher
variation in the conventional energy dispatched and thus a larger ramping cost. To reduce the ramping cost, the ISO needs to be more conservative in
dispatching the renewable energy, which results in a higher total cost. However, we can also see from the simulation that when the aggregators have
larger capacity to store energy, the increase of the total cost with the uncertainty is smaller. This is because the energy storage can smooth the
demand, in order to mitigate the impact of uncertainty in the renewable energy generation. This shows the value of energy storage to reduce the cost.

%

%

\subsubsection{Fairness}
\begin{figure}
\centering
\includegraphics[width =2.4in]{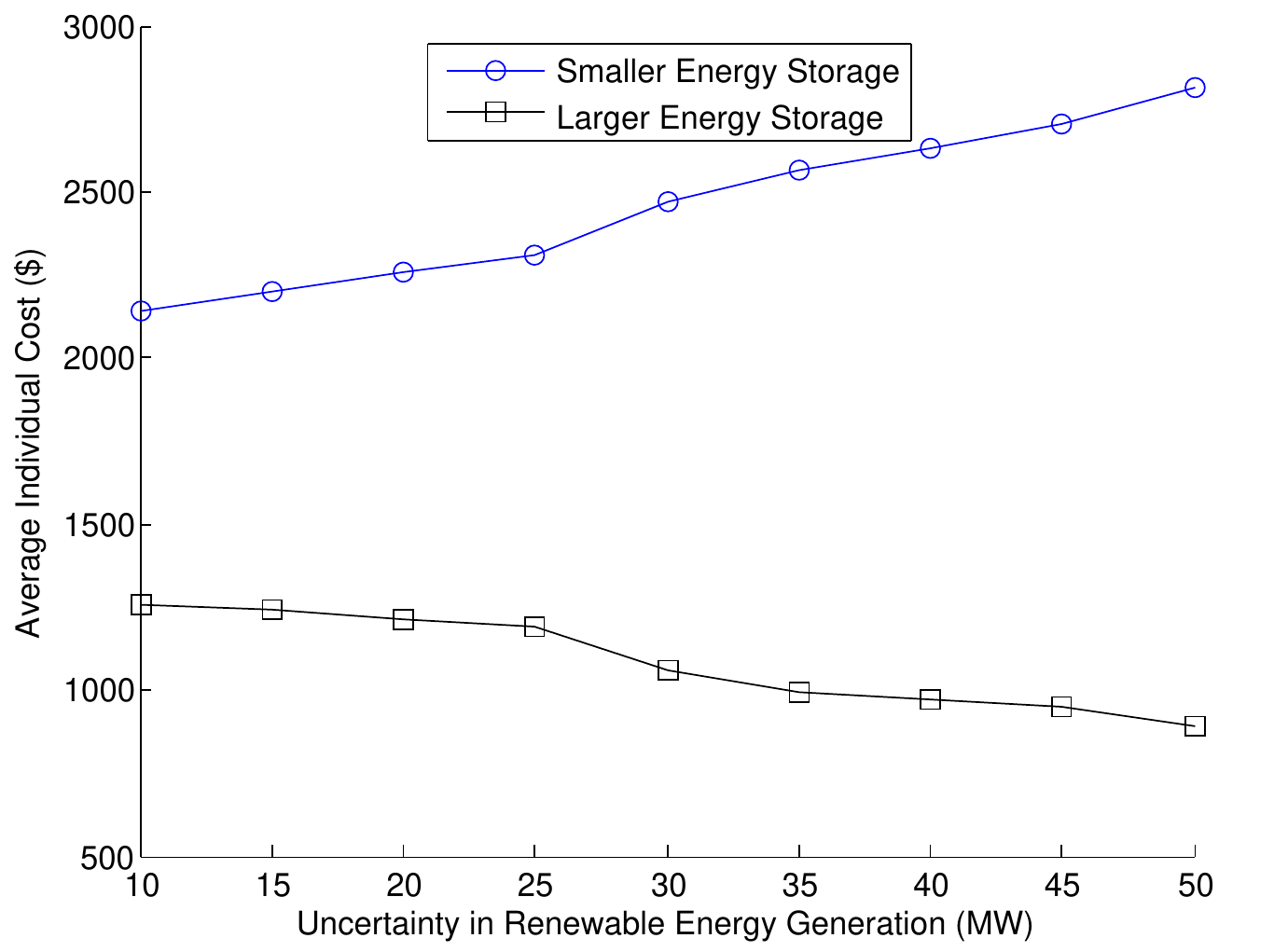}
\caption{The average individual costs of the aggregators with different energy storage in the IEEE 118-bus system.} \label{fig:Fairness}
\end{figure}

Now we investigate how the individual costs of the aggregators are influenced by the capacity of their energy storage. In particular, we are
interested in whether some aggregators are affected by having smaller energy storage. We assume that half of the aggregators have energy storage of
capacity 50~MW, while the other half have energy storage of much smaller capacity 10~MW. In Fig.~\ref{fig:Fairness}, we compare the average
individual cost of the aggregators with smaller energy storage and that of the aggregators with larger energy storage. We can see that the average
cost of the aggregators with smaller energy storage does increase with the uncertainty in renewable energy generation. Hence, the aggregators with
higher energy storage have an advantage over those with smaller energy storage, because they have high flexibility in coping with the price
fluctuation.

\subsection{The Conjectured Prices}
We compare the conjectured prices with the expected real prices. In our simulation, each aggregator $i$'s conjectured price is the conjectured price
that the proposed algorithm in Table~\ref{table:DecentralizedStrategy} converges to, namely $\lim_{k\rightarrow\infty} \tilde{y}_i^{(k)}$. We can
also calculate the expected real price as follows. The optimal DSM strategy that the algorithm in Table~\ref{table:DecentralizedStrategy} converges
to will induce a probability distribution over the states. In each state, we calculate the locational marginal price (LMP) for each aggregator based
on the actions taken at this state. Then we calculate the expected LMP of each aggregator, which is the expected real price at which each aggregator
pays for the energy.

\begin{figure}
\centering
\includegraphics[width =2.4in]{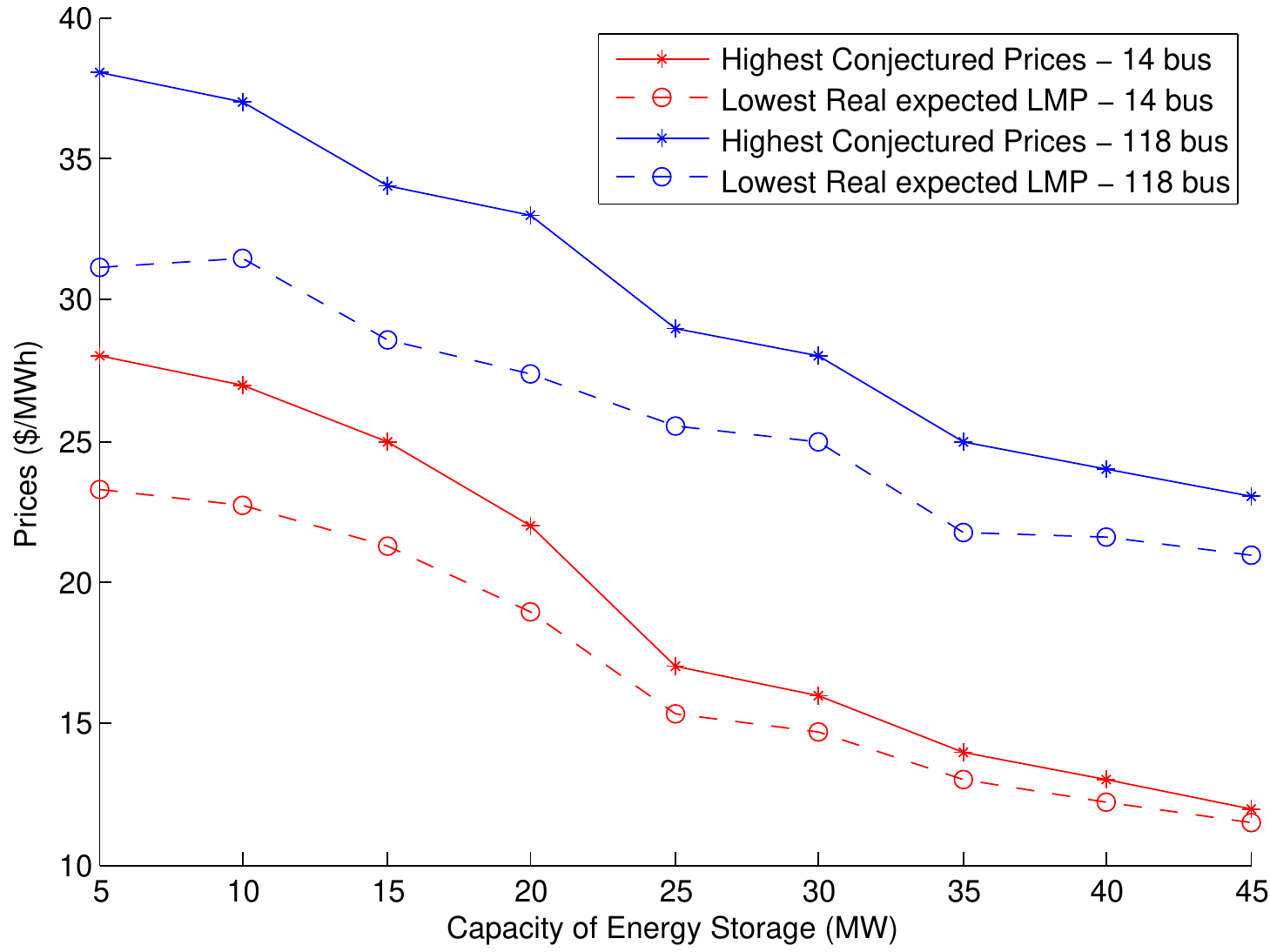}
\caption{The conjectured prices and the real expected locational marginal prices (LMPs) in IEEE 14-bus and 118-bus systems.} \label{fig:Prices}
\end{figure}

In Fig.~\ref{fig:Prices}, we show the highest conjectured price among all the aggregators and the lowest expected real price among all the
aggregators, because it is infeasible to plot the prices for all the aggregators in the figure. Hence, the difference of the conjectured price and
the real expected price for each individual aggregator is no larger than the difference shown in the figure. First, as we can see from the figure,
the prices go down when the capacity of the energy storage increases. This is because with energy storage, the congestion due to high energy purchase
demand decreases, which in turn decreases the congestion cost (technically, the Lagrangian multiplier associated with the capacity constraint is
smaller). Second, in our simulations (not shown in the figure), we observe that the conjectured prices are always higher than the expected real
prices. Hence, the conjectured price gives each aggregator an overestimate of the real expected price. An overestimate is better than an
underestimate, in the sense that each aggregator has a guarantee of how much it will pay in the worst case. Finally, we can see from the figure that
the conjectured price is close to the real expected price. The maximum difference between these two prices is less than 20\% in all the considered
scenarios.

\subsection{Comparisons of the DSM strategies}
In this section, to better understand why the proposed DSM strategies outperform the other strategies, we present a simple example and illustrate why
the proposed strategy achieves a lower cost. To keep the illustration simple, we reduce the number of states. Specifically, we assume that the demand
has three states: ``high'', ``medium'', ``low'', which corresponds to the highest, medium, and lowest values of the uniform distribution described in
\eqref{eqn:Simulation_DemandMeanValue}\eqref{eqn:Simulation_DemandDeviationFromMean}. Similarly, the energy storage has three states: ``empty'',
``half'', and ``full''. The maximum capacity of the renewable energy generator has two values, corresponding to the highest and lowest values in the
uniform distribution described in the basic simulation setup at the beginning of Section~\ref{sec:Simulation}. To distinguish the description of the
demand state, we suppose that the renewable energy generator harnesses solar energy, and refer to its states as ``sunny'' and ``cloudy'' instead of
``high'' and ``low''. We do not assume any randomness in the transmission lines. The purchase from each aggregator is also quantized into three
levels: large, moderate, and small.

In Table~\ref{table:Comparison_Strategies}, we compare the actions chosen by different strategies under different states in an IEEE 14-bus system.
Due to space limitation, we cannot show the strategies of all the aggregators, but only one of them. The state is a three tuple that consists of the
demand, the energy storage, and the renewable energy generation capacity. Although we have reduced the number of states, there are still
$3\times3\times2=18$ states, which is hard to show in one table. Instead, we only show the actions in some representative states, in which different
strategies take very different actions.

First, we can observe that the myopic strategy takes actions based on the demand and the energy storage exclusively. The myopic strategy aims to
minimize the current operational cost of the energy storage as long as the demand can be satisfied. Hence, it chooses to purchase small amount of
energy as long as the demand can be fulfilled by the energy left in the storage. Second, as we discussed at the end of
Section~\ref{sec:DesignFramework}, the strategy based on Lyapunov optimization does not take into account the demand dynamics. As we can see from the
table, the strategy based on Lyapunov optimization takes actions based on the energy storage and the renewable energy generation capacity
exclusively. It will purchase large amount of energy as long as it is sunny (which means that the capacity of the renewable energy generator is high
and hence the price is low). In contrast, the proposed strategy considers all the three states when making decisions. For example, when the states
are (low,empty,sunny) and (high,empty,sunny), the strategy based on Lyapunov optimization always chooses to purchase large amount of energy, while
the proposed strategy will purchase moderate amount of energy when the demand is low. Finally, the strategy based on MU-MDP also considers all the
three states, and takes similar actions as the proposed strategy. However, the strategy based on MU-MDP takes more conservative actions (e.g.
purchases small amount of energy when the proposed strategy purchases moderate amount of energy). This is because there is only one Lagrangian
multiplier under all the states, and to ensure the feasibility of the constraints, the Lagrangian multiplier has to be set larger. This results in a
harsher penalty in the objective function. Hence, the actions taken are more conservative to ensure that the line capacity constraints are satisfied.

\begin{table}
\renewcommand{\arraystretch}{1.1}
\caption{Comparisons of different strategies.} \label{table:Comparison_Strategies} \centering
\begin{tabular}{|c|c|c|c|c|c|c|}
\hline
State & (high,full,sunny) & (high,full,cloudy) & (low,full,sunny) & (low,empty,sunny) & (high,empty,cloudy) & (high,empty,sunny) \\
\hline
Myopic & small & small & small & small & large & large \\
\hline
Lyapunov & large & small & large & large & small & large \\
\hline
MU-MDP & large & small & small & small & small & large \\
\hline
Proposed & large & low & moderate & moderate & low & large \\
\hline
\end{tabular}
\end{table}


\section{Conclusion} \label{sec:Conclusion}
In this paper, we proposed a methodology to perform optimal foresighted DSM strategies that minimize the long-term total cost of the power system. We
overcame the hurdles of information decentralization and complicated coupling in the system, by decoupling the entities' decision problems using
conjectured prices. We proposed an online algorithm for the ISO to update the conjectured prices, such that the conjectured prices can converge to
the optimal ones, based on which the entities make optimal decisions that minimize the long-term total cost. We prove that the proposed method can
achieve the social optimum, and demonstrate through simulations that the proposed foresighted DSM significantly reduces the total cost compared to
the optimal myopic DSM (up to 60\% reduction), and the foresighted DSM based on the Lyapunov optimization framework (up to 30\% reduction).

\appendices
\section{Proof of Theorem~\ref{theorem:Convergence}}\label{proof:Convergence}

Due to limited space, we give a detailed proof sketch. The proof consists of three key steps. First, we prove that by penalizing the constraints
$\bm{f}(s_0,a_0,\bm{a})$ into the objective function, the decision problems of different entities can be decentralized. Hence, we can derive optimal
decentralized strategies for different entities under given Lagrangian multipliers. Then we prove that the update of Lagrangian multipliers converges
to the optimal ones under which there is no duality gap between the primal problem and the dual problem, due to the convexity assumptions made on the
cost functions. Finally, we validate the calculation of the conjectured prices.

First, suppose that there is a central controller that knows everything about the system. Then the optimal strategy to the design problem
\eqref{eqn:DesignProblem} should result in a value function $V^*$ that satisfies the following Bellman equation: for all $s_0,s_1,\ldots,s_I$, we
have
\begin{eqnarray}\label{eqn:Bellman_Original}
V^*(s_0,\bm{s}) = & \max_{a_0,\bm{a}} & \left\{ (1-\delta) \cdot \sum_{i=0}^I c_i(s_i,a_i) + \delta \cdot \sum_{s_0^\prime,\bm{s}^\prime}
\rho(s_0^\prime,\bm{s}^\prime|s_0,\bm{s},a_0,\bm{a}) V^*(s_0^\prime,\bm{s}^\prime) \right\} \\
& s.t. & \bm{f}(s_0,a_0,\bm{a}) \leq 0. \nonumber
\end{eqnarray}

Defining a Lagrangian multiplier $\bm{\lambda}(s_0)\in\mathbb{R}_+^N$ associated with the constraints $\bm{f}(s_0,a_0,\bm{a}) \leq 0$, and penalizing
the constraints on the objective function, we get the following Bellman equation:
\begin{eqnarray}\label{eqn:Bellman_Lambda}
V^{\bm{\lambda}}(s_0,\bm{s}) = & \max_{a_0,\bm{a}} & \left\{ (1-\delta) \cdot \left[\sum_{i=0}^I
c_i(s_i,a_i)+\bm{\lambda}^T(s_0)\cdot\bm{f}(s_0,a_0,\bm{a})\right] \right. \\
& & \left. + \delta \cdot \sum_{s_0^\prime,\bm{s}^\prime} \rho(s_0^\prime,\bm{s}^\prime|s_0,\bm{s},a_0,\bm{a})
V^{\bm{\lambda}}(s_0^\prime,\bm{s}^\prime) \right\}. \nonumber
\end{eqnarray}

In the following lemma, we can prove that \eqref{eqn:Bellman_Lambda} can be decomposed.

\begin{lemma}\label{lemma:DecompositionValueFunction}
The optimal value function $V^{\bm{\lambda}}$ that solves \eqref{eqn:Bellman_Lambda} can be decomposed as $V^{\bm{\lambda}}(s_0,\bm{s}) =
\sum_{i=0}^I V_i^{\bm{\lambda}}(s_i)$ for all $(s_0,\bm{s})$, where $V_i^{\bm{\lambda}}$ can be computed by entity $i$ locally by solving
\begin{eqnarray}
\begin{array}{c} V_i^{\bm{\lambda}(s_0)}(s_i) = \max_{a_i} \left\{ (1-\delta) \cdot \left[c_i(s_i,a_i)+\bm{\lambda}^T(s_0) \cdot f_i(s_0,a_i)\right] + \delta
\cdot \sum_{s_i^\prime} \rho_i(s_i^\prime|s_i,a_i) V_i^{\bm{\lambda}}(s_i^\prime) \right\} \end{array}.
\end{eqnarray}
\end{lemma}
\begin{IEEEproof}
This can be proved by the independence of different entities' states and by the decomposition of the constraints $\bm{f}(s_0,a_0,\bm{a})$.
Specifically, in a DC power flow model, the constraints $\bm{f}(s_0,a_0,\bm{a})$ are linear with respect to the actions $a_0,a_1,\ldots,a_I$. As a
result, we can decompose the constraints as $\bm{f}(s_0,a_0,\bm{a}) = \sum_{i=0}^I f_i(s_0,a_i)$.
\end{IEEEproof}

We have proved that by penalizing the constraints $\bm{f}(s_0,a_0,\bm{a})$ using Lagrangian multiplier $\bm{\lambda}(s_0)$, different entities can
compute the optimal value function $V_i^{\bm{\lambda}(s_0)}$ distributively. Due to the convexity assumptions on the cost functions, we can show that
the primal problem \eqref{eqn:DesignProblem} is convex. Hence, there is no duality gap. In other words, at the optimal Lagrangian multipliers
$\bm{\lambda}^*(s_0)$, the corresponding value function $V^{\bm{\lambda}^*(s_0)}(s_0,\bm{s}) = \sum_{i=0}^I V_i^{\bm{\lambda}^*(s_0)}(s_i)$ is equal
to the optimal value function $V^*(s_0,\bm{s})$ of the primal problem \eqref{eqn:Bellman_Original}. It is left to show that the update of Lagrangian
multipliers converge to the optimal ones. It is a well-known result in dynamic programming that $V^{\bm{\lambda}(s_0)}$ is convex and piecewise
linear in $\bm{\lambda}(s_0)$, and that the subgradient is $\bm{f}(s_0,a_0,\bm{a})$. Note that we use the sample mean of $a_0$ and $\bm{a}$, whose
expectation is the true mean value of $a_0$ and $\bm{a}$. Since $\bm{f}(s_0,a_0,\bm{a})$ is linear in $a_0$ and $\bm{a}$, the subgradient calculated
based on the sample mean has the same mean value as the subgradient calculated based on the true mean values. In other words, the update is a
stochastic subgradient descent method. It is well-known that when the stepsize $\Delta(k)=\frac{1}{k+1}$, the stochastic subgradient descent will
converge to the optimal $\bm{\lambda}^*$.

Finally, we can write the conjectured prices by taking the derivatives of the penalty terms. For aggregator $i$, its penalty is $\bm{\lambda}^T(s_0)
\cdot f_i(s_0,a_i)$. Hence, its conjectured price is
\begin{eqnarray}
\begin{array}{c} \frac{\partial \bm{\lambda}^T(s_0) \cdot f_i(s_0,a_i)}{\partial a_i} = \bm{\lambda}^T(s_0) \cdot \frac{\partial f_i(s_0,a_i)}{\partial a_i} \end{array}.
\end{eqnarray}

\end{document}